\documentclass[a4paper,11pt]{article}
\usepackage[skip=0pt]{caption}
\usepackage{jheppub} % for details on the use of the package, please
\usepackage[T1]{fontenc}
\usepackage{multirow}
\usepackage{longtable}
\usepackage{graphicx}% Include figure files
\usepackage[export]{adjustbox}% alignment for figure files
\usepackage{dcolumn}% Align table columns on decimal point
\usepackage{bm}% bold math
\usepackage{epsfig}
\usepackage{gensymb}
\usepackage{mathrsfs}
\usepackage{extpfeil}
\usepackage{extarrows}
\usepackage{float}
\usepackage{pdfpages}
\usepackage{amsmath}
\usepackage{amssymb}
\usepackage{graphicx,bm}
\usepackage{natbib}
\usepackage{slashed}
\usepackage[makeroom]{cancel}
\usepackage{dsfont}
\usepackage{longtable}
\usepackage{multirow}
\usepackage{youngtab}
\usepackage{young}
\usepackage{tikz}
\newcommand{\scr}[1]{\ensuremath{\mathcal{#1}}}
\newcommand{\nn}{\nonumber\\}
\def\[#1\]{\begin{align}#1\end{align}}
\newcommand{\eql}[1]{\label{eq:#1}}
\newcommand{\Eq}[1]{Eq.~\eq{#1}}
\newcommand{\Eqs}[1]{Eqs.~\eq{#1}}
\newcommand{\eq}[1]{(\ref{eq:#1})}  % equation reference without "Eq."
\newcommand{\ggap}{\hspace{0.1em}}% useful for adjusting space in formulas

\def\o{{\textrm{O}}}
\def\e{{\textrm{E}}}
\def\b{\textrm{B}}

 \usepackage[titletoc]{appendix}

\def\fun#1#2{\lower3.6pt\vbox{\baselineskip0pt\lineskip.9pt
  \ialign{$\mathsurround=0pt#1\hfil##\hfil$\crcr#2\crcr\sim\crcr}}}

\usepackage{tikz}
\usetikzlibrary{decorations.pathmorphing,decorations.markings,trees,shapes}
\usepackage[compat=1.0.0]{tikz-feynman}

\def\lsim{\mathrel{\rlap{\raise 2.5pt \hbox{$<$}}\lower 2.5pt\hbox{$\sim$}}}
\def\gsim{\mathrel{\rlap{\raise 2.5pt \hbox{$>$}}\lower 2.5pt\hbox{$\sim$}}}

\input epsf

\usepackage{color}

\newcommand{\comment}[1]{}

\newcommand{\be}{\begin{equation}}
\newcommand{\ee}{\end{equation}}
\newcommand{\bea}{\begin{eqnarray}}
\newcommand{\eea}{\end{eqnarray}}

\makeatletter
\newcommand{\hathat}[1]{% 
\begingroup%
  \let\macc@kerna\z@%
  \let\macc@kernb\z@%
  \let\macc@nucleus\@empty%
  \widehat{\raisebox{.4ex}{\vphantom{\ensuremath{#1}}}\smash{\widehat{#1}}}%
\endgroup%
}
\makeatother

\renewcommand\arraystretch{2}
\makeatletter
\gdef\@fpheader{}
\makeatother

\begin{document}

\title{Quantifying EFT Uncertainties in LHC Searches}

\author[1]{Spencer Chang}
\author[2]{\!\!, Markus A.~Luty}
\author[3]{\!\!, Teng Ma}
\author[4]{\!\!, Francesco Montagno}
\author[4,5]{\!\! and Andrea Wulzer}
\affiliation[1]{Department of Physics and Institute for Fundamental Science,\\
University of Oregon, Eugene, Oregon 97403,  USA}
\affiliation[2]{Center for Quantum Mathematics and Physics (QMAP),\\ University of California, Davis, California 95616}
\affiliation[3]{
International Center for Theoretical Physics Asia-Pacific (ICTP-AP),\\
University of Chinese Academy of Sciences, 100190 Beijing, China}
\affiliation[4]{Institut de F\'{\i}sica d'Altes Energies (IFAE), The Barcelona Institute of Science and Technology (BIST),
Campus UAB, 08193 Bellaterra, Barcelona, Spain}
\affiliation[5]{ICREA, Instituci\'o Catalana de Recerca i Estudis Avan\c{c}ats, 
Passeig de Llu\'{\i}s Companys 23, 
08010 Barcelona, Spain}

\abstract{
Effective Field Theory (EFT) is a general framework to parametrize the low-energy approximation to a UV model 
that 
is widely used  in model-independent searches for new physics. 
The use of EFTs at the LHC can suffer from a `validity' issue, since new physics amplitudes often grow with energy and the kinematic regions with the most sensitivity to new physics have the largest theoretical uncertainties.
We propose a method to account for these uncertainties with the aim of producing robust model-independent results with a well-defined statistical interpretation.  In this approach, one must specify the new operators being studied as well as the new physics cutoff $M$, the energy scale where the EFT approximation breaks down.  
At energies below $M$, the EFT uncertainties are accounted for by adding additional higher dimensional operators with coefficients that are treated as nuisance parameters.  
The size of the nuisances are governed by a prior likelihood function that incorporates information about dimensional analysis, naturalness, and the scale $M$. 
At energies above $M$, our method incorporates the lack of predictivity of the EFT, and we show that 
this
is crucial to obtain consistent results. 
We perform a number of tests of this method in a simple toy model, illustrating its performance in analyses aimed at new physics exclusion as well as for discovery.  
The method is conveniently implemented by the technique of event reweighting and is easily ported to realistic LHC analyses.  
We
find that the procedure converges quickly with the number of nuisance parameters and is conservative when compared to UV models.  
The paper gives a precise meaning and offers a principled and practical solution to the 
widely debated `EFT validity issue'.
}

\maketitle

% =============================================================================
% =============================================================================
% =============================================================================
% =============================================================================
\section{Introduction}

All theoretical predictions rely on approximations and any comparison between theory and experiment is subject to uncertainties in the theoretical predictions.
Scientific inferences drawn from comparing theory and data is therefore unavoidably based on assumptions on the magnitude of the theory uncertainties.
This paper aims at making these assumptions explicit and precise for the theory uncertainties that arise when we approximate an unknown underlying UV-complete
theory with a low-energy Effective Field Theory (EFT).
The problem is similar to the estimate of missing higher orders in a perturbative expansion, but 
there are important methodological and conceptual differences.

An EFT is an approximate description of a particle physics model 
that includes only the particles that are light enough to be directly produced in experiments. 
The effect of additional heavy particles is included indirectly through local interactions with coupling strengths known as Wilson coefficients. 
In this paper we focus on the case where the light particles are those 
of the Standard Model (SM) and the heavy particles are unknown new particles
above the electroweak scale.

If the underlying UV model is known, the EFT can be derived straightforwardly and the 
Wilson coefficients computed in terms of the parameters of the model. 
In this case, the EFT is simply a tool to simplify calculations. 
However, EFT is especially useful when the underlying model is not completely
specified, because the same 
set of EFT interactions emerge from a large class of different UV models. 
By treating the Wilson coefficients as free parameters and comparing with the data we 
are thus effectively probing many different UV models at once. 
In fact, the correct model that describes the data may be completely unknown,
and still deviations from the SM can be probed in a general `model-independent' 
EFT search. 
For this reason, there is widespread agreement that precision measurements at the LHC are naturally interpreted as measurements or constraints on EFT Wilson coefficients, for example in the SMEFT framework~\cite{Buchmuller:1985jz,Grzadkowski:2010es}. 
However, there is currently no agreed-upon method for quantitatively estimating the
theory uncertainties arising from the approximate nature of the EFT \cite{Brivio:2022pyi}.  
EFTs are also useful in a variety of different contexts beyond searches at the LHC,
providing additional motivation for the development of a principled paradigm for 
the estimation of the EFT uncertainties.

Estimating the theoretical uncertainties from an EFT poses novel conceptual issues because
the EFT is an inherently \emph{incomplete} theoretical description.
An EFT describes the effects of heavy particles with mass $M$ in an
expansion in the small parameter $E/M$, where $E$ is the energy scale of a
physical process.
For $E \ll M$, the EFT uncertainties can be parameterized by 
a series of interaction terms whose effects are suppressed by powers of $E/M$.
However, the EFT loses all predictive power for $E \gsim M$,
where the heavy particles can be produced.
In this regime, the scattering amplitude becomes 
a non-analytic function of the 
kinematical variables and can no longer be expanded in 
powers of $E/M$.
If the EFT predictions are extrapolated into this regime, they generally
violate basic theoretical principles, such as unitarity, and therefore do not
resemble any possible UV-complete model.
These questions are of immediate practical relevance at the LHC, where the 
combination of high
energy and limited precision means
that the heavy particle scales $M$ that are probed in experiments 
are often
not far from 
the energy scale of the processes of interest.
In particular, one must ensure that data in this high-energy 
region is not used to constrain the EFT.
This is the so-called `EFT validity' issue.

This problem manifests itself concretely in cases where the EFT leads to amplitudes that grow with energy.
For example, there are a number of 2-to-2 scattering amplitudes in dimension-6 SMEFT that
grow with energy as $G E^2$ at energies $E$ above the masses of SM particles,
where $G$ is a Wilson coefficient of a dimension-6 operator.
This energy-growing
behavior is beneficial for the sensitivity to new physics,
since the SM 
contribution to the scattering amplitude
does not have this energy growth.
On the other hand, in complete models this energy growth is cut off at the
scale $M$ of new heavy particles,
so extrapolating the EFT into this regime gives incorrect results.

Our approach to this problem is to introduce parameterized uncertainties into the  EFT prediction that reflect the higher order corrections and
limited regime of validity of the EFT.
This is done by defining a modified amplitude that depends on a number of
nuisance parameters that parameterize the uncertainty of the EFT prediction.
The EFT uncertainties depend on the
scale $M$ at which the EFT breaks down,
which is included  as an independent parameter in the analysis.
The constraints on the Wilson coefficients therefore depend on the 
assumed value of the cutoff $M$.
For events with energy $E \ll M$, the parameterized uncertainties are small, while
for $E \gsim M$ the uncertainties are large, reflecting the predictive power of an EFT 
with cutoff $M$. 
This solves the EFT validity issue,
since data in kinematic regimes where the EFT is invalid do not constrain
the Wilson coefficients and are therefore effectively excluded from the
analysis.
Furthermore, the method provides a quantitative estimate of the impact of
unknown higher order EFT operators at low energies.

This is similar in spirit to the treatment of theory uncertainties from missing higher orders in the loop expansion, 
for instance in QCD. 
Also in that case, one parametrizes the unknown contributions using nuisance parameters based on an estimate of their size.
(For a recent discussion, see \cite{Tackmann:2024kci}.) 
The conceptual difference is that missing higher orders could in principle be computed, while the EFT uncertainties cannot be computed because the microscopic theory is not known.

The EFT uncertainties are modeled using the following logic.
We want the truncated EFT with cutoff $M$ to describe the 
low-energy physics of a large class of UV models with new physics at the scale $M$.
The EFT uncertainty arises from the fact that we do not know the correct UV model.
We treat this uncertainty statistically by parameterizing
these unknown effects in the truncated EFT predictions with nuisance parameters.
This means that the constraints obtained depend on prior assumptions
about the size of the allowed deviations from the truncated EFT.
We stress that this is an unavoidable feature of any EFT search,
since a truncated EFT has no predictive power without some assumptions
about the size of the omitted terms.
Our method makes these assumptions explicit and incorporates them into
a well-defined statistical framework.
We believe that our assumptions are reasonable, 
but it is also straightforward to modify them
to aid in interpreting the results.

Fig.~\ref{fig:uv_reach_with_clipping} 
shows the results of our method and
illustrates the EFT validity issue in a toy setup for EFT exclusion of a Wilson coefficient
$G$,  defined in \S\ref{subExc}. 
Note that the EFT search should give exclusion constraints 
that are \emph{weaker} than
the constraints for any UV model that predicts the same value of $G$,
because the EFT is an incomplete description with limited predictive power.
Instead, we show that the exclusion reach of a `plain EFT' analysis 
that includes no EFT uncertainties
(black dashed line)
is \emph{stronger} than the reach
obtained in a UV model where the EFT operator is generated by the exchange 
of a new heavy resonance of mass $M$ in the $t$-channel ($G_{\textrm{exc}}^{(t)}$, solid black curve). 
This is because the EFT analysis assumes  $G E^2$ corrections to the SM amplitude at all energies, which are not present 
in the UV model,
resulting in 
a too-strong exclusion. 
This is clearly an artifact of the inconsistent usage of the EFT beyond the cutoff.
The discrepancy becomes small only for $M \gsim 5$~TeV, meaning that the result of a plain EFT search 
cannot be interpreted as a constraint on a Wilson coefficient generated by typical UV models
without an assumption of the value of $M$.
However, if one includes the EFT uncertainties using our method, it 
produces an exclusion (orange curve) that is weaker than the UV-model limit for any $M$,
as expected for a model-independent result.

Our proposal has similarities with a number of works in the literature that estimate uncertainties in constraints on dimension-6 SMEFT operators by including the effects of dimension-8 operators.
In particular, Refs.~\cite{Boughezal:2022nof,Allwicher:2022gkm} are most similar to our approach.
However, we believe that our method has significant advantages over these approaches,
and we will compare our method to theirs in \S\ref{sec:comparison}.

A
strategy for dealing with the limited range of validity of EFTs that has been adopted at ATLAS and CMS
is to select
low-energy data for which the EFT is a good approximation. For EFT searches at the LHC, this means cutting on a measured observable that is representative of the partonic scattering energy. For example, one can use transverse momentum $p_T$ of some hard central object and exclude from the analysis high energy data with $p_T > M_\text{clip}$. The `clipping scale' $M_\text{clip}$ is a floating parameter that is qualitatively linked to the EFT cutoff $M$.  In such analyses, the constraints
are presented as a function of $M_\text{clip}$, thus displaying the sensitivity of the results to data at different energy scales. 
Provided that the observable ($p_T$ in our example) is truly representative of the hardness of the underlying partonic event, data clipping succeeds in avoiding the use of data in the kinematic regime where the EFT is completely unreliable.

However, data clipping has a number of problems. 
First, it is not always possible to identify an experimental variable that faithfully represents the hard scattering scale. 
For example, events with $p_T < M_\text{clip}$ can arise from partonic events with large center of mass energy, so a $p_T$ cut does not necessarily guarantee that the event is accurately described by the EFT. 
In general, even in final states where a better variable than $p_T$ can be measured, the selected data can still suffer  contamination from high-energy 
events
due to the difficulty of isolating the products of the hard collision from the large flux of energy in the forward region, coming from the underlying event or initial state radiation. The impact of the contamination should be quantified, but there is no strategy for this in the clipping method. 
Moreover, the choice of the clipping variable depends on the 
scattering process used in the analysis,
while a universal process-independent methodology
is needed to combine
EFT searches in different final states, as is required for a comprehensive global EFT fit. In our approach it is instead possible to combine searches as the cutoff $M$ and nuisances are   
simply universal parameters in the model being constrained.  

The second limitation of the data clipping approach is the qualitative nature of the connection between the clipping scale and the EFT cutoff scale. Identifying the two scales is not correct, because the clipping method treats the EFT predictions as perfect if the energy is only slightly below $M_\text{clip}$, without accounting for the EFT uncertainties that must become large as the energy approaches the cutoff $M$. 
This is clearly not a good model of the EFT uncertainties and Fig.~\ref{fig:uv_reach_with_clipping} shows that the bounds obtained from clipped searches can be stronger
than the bounds for UV models if we identify $M_\text{clip}$ with the mass of the heavy resonances.
This means that $M_\text{clip}$ should be taken somewhat below $M$, so that EFT uncertainties are suppressed enough to be neglected, but there is no clear way to go beyond this qualitative observation. 
This means that we cannot interpret a limit from a search with clipping scale $M_\text{clip}$ as a bound on models where the EFT operators are generated by the exchange of heavy particles with mass near $M_\text{clip}$.
Our proposal solves these problems by introducing uncertainties into
the model prediction in a way that reflects
the effects of unknown new particles at the scale $M$.

This paper is organized as follows. 
In \S\ref{sec:howto} we explain 
our parameterization of the EFT uncertainties and the physical ideas it is based on.
In the region of energy below $M$, we enforce the theoretical knowledge that the corrections are small and controlled by powers of $E/M$ up to order one factors. In the region above $M$, our task is instead to enforce theoretical ignorance by replacing the EFT predictions with a flexible
parametrization of the scattering amplitude. Practical considerations are also taken into account: we formulate our proposal in a way that the dependence on the nuisance parameters of the predictions can be included through the technique of event-reweighting.
This allows the use of
standard simulation tools that are fully integrated in the ATLAS and CMS simulation chain,
without requiring more simulations than needed for a regular EFT analysis.  
At the end of this section, we preview
our results and  compare with some existing proposals in the literature. 

In \S\ref{sec:CI} we describe the implementation of our modeling of the EFT uncertainties in an EFT search aimed at exclusion. 
We work in a toy setup, which is however realistic enough to demonstrate its portability to a real LHC analysis. 
We also investigate in detail the role
played in the analysis by the different elements of our proposal, such as the nuisance parameters with their prior, as well as the form factors used to mitigate the energy growth of the EFT predictions. We will assess the robustness of the results under significant variations of these elements with respect to our baseline prescription.

In \S\ref{sec:Res} we show the results our method, both for exclusion and for new physics discovery, and check their consistency with searches for explicit models. 
In particular, we verify that our approach robustly solves the EFT validity issue as previously discussed. 
We also find that our method can have a better reach for discovery than the 
`plain EFT' analysis despite
the fact that it introduces uncertainties, because it gives a more accurate representation of the new physics effects at higher energy.

Our conclusions are reported in \S\ref{sec:conc}. 
The paper contains three appendices with supplementary discussion.
Appendix~\ref{app:TP} addresses the parametrization of uncertainties for 3-point 
EFT interactions. 
Appendix~\ref{app:MO} describes the straightforward generalization of the implementation when more than one EFT operator is considered simultaneously. 
In Appendix~\ref{subcorr} we critically discuss the large-sample and Asimov approximations 
used in the statistical analyses in the paper.

\section{Parameterizing EFT Uncertainties}
\label{sec:howto}

An EFT provides an approximate low-energy description of a large class of UV models. The EFT is universal in the sense that the allowed interactions in the EFT depend only on the light particle content and the symmetries of the theory.
However, an EFT contains infinitely many terms and in order to be predictive, it must be truncated to a finite 
subset.
Our starting point is that a truncated EFT of interest has been identified---we refer to it as `the EFT' for brevity---and that we want to perform an experimental search to constrain its Wilson coefficients. 
In this paper, we focus on the case where the only light particles are
those of the SM, so the new particles are above the electroweak scale.
This means that we are studying the Lagrangian
\[
\eql{EFTtrunc1}
\scr{L}=\scr{L}_\text{SM}
+ \sum\limits_{\scr{O}} G_{\scr{O}} \, \scr{O},
\]
where $\scr{O}$ are additional local operators added to the SM Lagrangian
and $G_{\scr{O}}$ are the Wilson coefficients we want to probe.
We will focus on experiments, at or above the electroweak scale $\sim 100$~GeV, and operators $\scr{O}$ parameterizing 2-to-2 processes among the SM particles, although our method has potential applications beyond this. The generalization of our methodology to include
3-point couplings is discussed in Appendix~\ref{app:TP}.

The criteria that are employed for the selection of the operators of interest are not relevant for the present discussion. 
One could start for instance from the whole set of dimension-6
SMEFT operators compatible with flavor symmetries and restrict to the interactions that can be probed in the LHC process of interest.
A more principled approach is to apply a `power-counting rule' that dictates the relative importance of the different terms. 
Power counting is inherited from the UV realization of the EFT and thus the correct truncation depends on what type of UV physics the EFT is describing. For example, if the Higgs is a composite Nambu--Goldstone boson at high energies, the leading terms in the EFT are given by the SILH framework~\cite{Giudice:2007fh}.
Our method is designed to apply to these, or any other EFT truncation.

\subsection{Mandelstam Descendants}
The EFT gives an approximation of 
scattering amplitudes
with characteristic energy $E$ smaller than some scale $M$ (called the cutoff), which is generally the mass of the lightest BSM particle in a complete
description.
In the regime $E \ll M$, the EFT 
is essentially an expansion of the amplitude in powers of the small parameter $E/M$.
 
In the complete theory, the leading terms in the EFT are obtained by 
approximating the propagator of heavy particles as a constant and there are an
infinite number of higher order corrections coming from the expansion
of the propagator in powers of $E/M$.
For example, the 2-to-2 fermion scattering amplitude mediated by tree-level exchange of a heavy scalar
with mass $M$ and Yukawa coupling $y$ in the $s$-channel can generate
an effective operator $(\bar\psi \psi)^2$ in the EFT.
The amplitude is given by 
\bea  
\!\!\!\!\!\!\!
\scr{M}_{\text{BSM}}(12 \to 34)
&=& 
 \includegraphics[valign=c,scale=0.35]{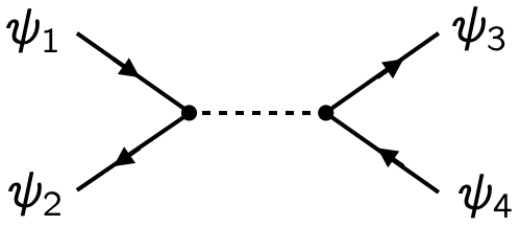}
\nonumber  \\
&=& G_{(\bar\psi \psi)^2} 
\biggl[ 1 + \frac{\hat{s}}{M^2} 
+ \frac{\hat{s}^2}{M^4}  + \cdots \biggr],
\qquad
G_{(\bar\psi \psi)^2} = \frac{y^2}{M^2}\,,
\eql{Mtree}
\eea 
where the second line has the low energy expansion in $\hat{s}$, the Mandelstam $s$ variable of the partonic scattering process. The exchange of resonances in other channels would similarly produce polynomial corrections in the other Mandelstam variables $\hat{t}$ and $\hat{u}$.

The same pattern of corrections is encountered if the operator 
is generated
by integrating out a heavy scalar with mass $M$ at one loop:
\bea 
&&\!\!\!\!\!\!\!
\scr{M}_{\text{BSM}}(12 \to 34)
= \includegraphics[valign=c,scale=0.35]{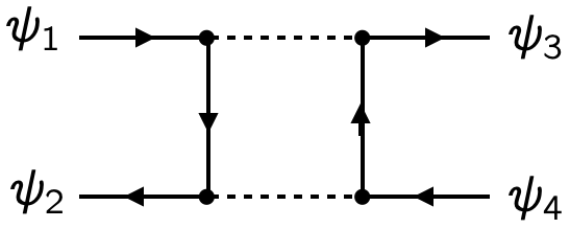}
\nonumber \\
&&= G_{(\bar\psi \psi)^2}
\biggl[ 1 + c_1 \frac{\hat{s}}{M^2} + c_2 \frac{\hat{t}}{M^2}
+ c_3 \frac{\hat{s}^2}{M^4} + c_4 \frac{\hat{s}\,\hat{t}}{M^4} +
\cdots \biggr] + \cdots,
\eql{Mloop}
\eea 
where `$\cdots$' denotes additional operators generated by the loop, and
\bea
G_{(\bar\psi \psi)^2} \sim \frac{y^4}{16\pi^2 M^2}\, .
\eea 
The loop factor and the coupling $y$ of the complete UV theory are absorbed into the expression for the Wilson coefficient $G_{(\bar\psi \psi)^2}$.
The higher order terms now depend on all of the Mandelstam variables and their coefficients $c_1, c_2, \ldots$ defined 
above 
are of order one.

In \Eqs{Mtree} and \eq{Mloop}, the terms in brackets correspond to
additional EFT interactions with two or more derivatives.
We call these terms `Mandelstam descendants' of the original interaction and we will use them as the starting point for estimating the uncertainties
due to the incompleteness of the truncated EFT. 
The general structure we expect from integrating out heavy physics is
\[
\scr{M}_{\text{BSM}}(12 &\to 34)
=  \sum_{\scr{O}} G_{\scr{O}} \scr{M_{\scr{O}}}
\bigg[ 
1 + c_{\scr{O},1} \frac{\hat{s}}{M^2}
+ c_{\scr{O},2} \frac{\hat{t}}{M^2}
+ c_{\scr{O},3} \frac{\hat{s}^2}{M^4}
+ \cdots
\bigg],
\eql{genM}
\]
where $\scr{M}_{\scr{O}}$ is the amplitude arising from the addition of the operator $\scr{O}$ in the Lagrangian  and the higher order terms come from expanding the heavy propagators, as in the examples above. 

If the operator $\scr{O}$ arises
from a heavy particle of mass $M$, either at tree or loop level, the coefficients $c_{\scr{O}}$ in~\eq{genM} 
are at most order one, but may be much smaller in some UV models.
For example, 
in the case of a tree level exchange in the $s$-channel, $c_{\scr{O},1}$ and $c_{\scr{O},3}$ are of order one while $c_{\scr{O},2}$ is suppressed. All the $c_{\scr{O}}$'s are order one  if instead the operator is generated by two particles with similar mass $M$ that are exchanged in the $s$- and in the $t$-channel. We account for these different possibilities by 
using a prior likelihood (see below) that is maximal at $c_{\scr{O}} = 0$ and small for $c_{\scr{O}}$ much larger than 1.
This enforces the constraint that $c_{\scr{O}}$ cannot be much larger than 1, but may be much smaller.

In some 
UV models, the different operators $\scr{O}$ might be generated by particles with different mass, resulting in a different scale $M_{\scr{O}}$ 
governing 
the higher order terms for each operator. Eq.~\eq{genM} accounts for this possibility provided that $M$ is interpreted as the mass of the lightest heavy BSM particle (such that $M_{\scr{O}}\geq M$) and the prior likelihood allows the $c_{\scr{O}}$ parameters to be small. 
Since we are assuming that the EFT should capture a wide range of models, this approach is appropriately conservative and general.

In fact, the form of the amplitude in \Eq{genM} corresponds to a complete classification of EFT operators in terms of tree-level on-shell amplitudes, initiated in \cite{Shadmi:2018xan, Ma:2019gtx,Durieux:2019eor, Durieux:2020gip} and made systematic in \cite{Dong:2021vxo,Dong:2022mcv,Dong:2022jru,Liu:2023jbq, Chang:2022crb, Bradshaw:2023wco,Arzate:2023swz,Dong:2024dce}.
Operators that are not the descendants of any other operator are called `primary operators'.  
This is a useful concept because there are a finite number of primary operators in the SM with 4 or fewer external legs (given explicitly in \cite{Chang:2022crb, Bradshaw:2023wco,Arzate:2023swz}).
Since the descendants correspond to adding pairs of derivative operators to primary operators, the primaries and descendants have the same transformation properties under the symmetries of the SM.
Therefore, in `generic' UV models the primary operators give the leading deviations from the SM from new heavy physics, with the effect of the higher-derivative descendants suppressed by additional powers of $E/M$.

For this reason, we believe it is important to initiate a program of `bottom-up' phenomenology to probe a large number of these primary operators.
In addition,
there are physically interesting UV models where the  leading operators for some processes are not primary operators,
and it may be interesting to study
these in an EFT approach.
For example if the new physics comes from a sector where the Higgs is as a PNGB, the UV theory has approximate shift symmetries that suppress some primary operators compared to their descendants \cite{Liu:2016idz}.
It is straightforward to adapt our method to handle these cases: if the leading operator is a descendant, we treat its Wilson coefficient 
as the parameter of interest and include EFT uncertainties from further descendants using \Eq{genM}.

As stated earlier, the EFT approximation breaks down completely above the cutoff scale $M$ due to the non-analyticity from exchanged particles with mass $M$.
From a bottom-up perspective, 
we expect that $M$ must be below the scale of perturbative 
unitarity violation for the new BSM interactions.  
This imposes an upper bound on $M$ for fixed couplings $G_{\scr{O}}$.  
These unitarity bounds can be estimated simply and directly from the 
tree-level amplitudes
\cite{Abu-Ajamieh:2020yqi,Chang:2022crb} and are reported for the primary
operators in \cite{Chang:2022crb, Bradshaw:2023wco,Arzate:2023swz}.%
%
% \footnote{%
% We consider here the constraints on the hard amplitudes that stem from perturbative $S$-matrix unitarity. A new notion of `collider unitarity' was introduced in~\cite{Cohen:2021gdw}
% .}
%
\footnote{We consider here the constraints on the hard partonic amplitudes from $S$-matrix unitarity.
Ref.~\cite{Cohen:2021gdw} introduced a notion of `collider unitarity' and used this to define a test of validity for experimental EFT searches.}
In our framework, the bounds are easily included as independent theory constraints on the 
experimental parameter space $(M, G_{\cal O})$.
These bounds mean that for some operators, the values of the coupling $G_{\cal O}$ for which there is experimental sensitivity will necessarily have a low value of $M$ and therefore the EFT has large theoretical uncertainties. 
Taking these properly into account (as in the method we are proposing) is essential
to get reliable results in this situation. 

Our proposal is meant to be applicable in cases where the data contains events with partonic energy $E \gsim M$,
for which the EFT prediction is completely uncertain.
As discussed in the Introduction, the use of data clipping to avoid this regime has serious problems, so our method must provide predictions even for kinematics where the EFT description has completely broken down.
Our approach will be to parameterize the uncertainty in the EFT prediction for the amplitude by nuisance parameters.
These are defined to reproduce the higher order corrections in \Eq{genM} for $E \ll M$, and make the amplitude completely uncertain for $E \gsim M$.
This means that data with $E \gsim M$ cannot be used to constrain the EFT couplings, effectively removing these events from the analysis.
In the next section, we will present a simple parameterization of the amplitude with these features, and in later sections will will provide evidence that this is sufficient for many purposes.

\subsection{Nuisance Parameters and Form Factors from Rescaling}
\label{ss:formfactors}
Let us now give the details of our proposal. 
For definiteness we focus on a single EFT operator ${\scr{O}}$ with Wilson coefficient $G$. The generalization to include multiple operators is straightforward and is discussed in Appendix \ref{app:MO}. We write the parton amplitude being probed as
\[
\scr{M} = \scr{M}_\text{SM} +  G \scr{M}_{\scr{O}} + O(G^2).
\]
Here $\scr{M}$ is a 2-to-2 parton scattering amplitude and correspondingly $\scr{O}$ is an operator containing 4 physical SM fields ($h$, $W$, $Z$, {\it etc\/}).\footnote{The case where $\scr{O}$ is a 3-point coupling is discussed in Appendix~\ref{app:TP}.} Both $\scr{M}_\text{SM}$ and $\scr{M}_{\scr{O}}$ may contain radiative corrections.

In our approach, the EFT uncertainties are taken into account by modifying the BSM contribution $\scr{M}_{\scr{O}}$ by a multiplicative factor $\scr{R}_\nu$, which is equivalent to rescaling the Wilson coefficient in the amplitude as
\[
\eql{Rdefn}
G \to G \ggap \scr{R}_\nu(x_1, x_2, x_3),
\]
where
\[
x_1 = \frac{\hat{s}_{12}}{M^2},
\ \ 
x_2 = \frac{\hat{s}_{13}}{M^2},
\ \
x_3 = \frac{\hat{s}_{14}}{M^2}.
\]
Here we have introduced an arbitrary labeling of the external parton momenta $p_{1,2,3,4}$ with all momenta outgoing and defined the partonic Mandelstam invariants by
\[
\hat{s}_{ij} = (p_i + p_j)^2.
\]
We use this notation because the conventional Mandelstam variables $\hat{s}, \hat{t}, \hat{u}$ depend on the scattering channel;
for example, in the $12\to 34$ channel, we have $\hat{s} = \hat{s}_{12} = \hat{s}_{34}$, $\hat{t} = \hat{s}_{13} = \hat{s}_{24}$, and $\hat{u} = \hat{s}_{14}= \hat{s}_{23}$.
The Mandelstam variables are not independent, so the rescaling factor $\scr{R}_\nu$ could be expressed as a function of only two of them, but the general expression is more suited for our purposes. 
The rescaling factor ${\cal R}_\nu$ depends on a number of parameters $\nu$ to be treated as nuisance parameters, as described below.

Introducing the uncertainties as a kinematic-dependent rescaling of $G$ has an enormous practical advantage. In a Monte Carlo simulation, it is possible to access the dependence on $G$ of the weight of each event, as well as the truth-level partonic kinematics. Therefore, it is possible to include ${\cal R}_\nu$ in the theoretical predictions by event-reweighting~\cite{Gainer:2014bta,Frixione:2002ik,Alioli:2010xd,Frederix:2014hta,REPOLO,Mattelaer:2016gcx,Andreassen:2019nnm,Alwall:2014hca,Degrande:2020evl} using \Eq{Rdefn} without the need for
additional simulations beyond
those needed to compute the dependence of the predictions on $G$. This simulation workflow is explained in Section~\ref{sec:CI} in detail.

The rescaling factor $\scr{R}_\nu$ should reproduce the power series expansion in powers of 
the Mandelstam variables
at energies $E \ll M$, and should account for the lack of predictivity of the EFT in the regime $E \gsim M$.
To do this, we propose a rescaling factor of the form
\[
\label{eq:FFonlyR}
\scr{R}_\nu(x, y, z) = \left( \frac{F(x_\text{max})}{x_\text{max}} \right)^{{\mathrm p}/2} \!\!\!
\sum\limits_{\alpha, \beta, \gamma \,\ge\, 0}
\nu_{\alpha\beta\gamma} 
F(x_1)^\alpha F(x_2)^\beta F(x_3)^\gamma,
\]
where $x_\text{max} = \max\{ x_1, x_2, x_3\}$.
For each event, one of the $x_{1,2,3}$ will be positive and the others negative, so $x_\text{max}$ is proportional to the positive Mandelstam invariant that defines the scattering channel.
For example, for the channel $12 \to 34$, we have $x_\text{max} = x_1$.
\footnote{Event generators such as {\tt Madgraph} provide the channel information for each parton-level generated event, so that the variables $x_{1,2,3}$ can be uniquely identified.}
Up to an overall pre-factor, $\scr{R}_\nu$ is a general polynomial in 
$F(x_{1,2,3})$, 
with non-negative integer powers $\alpha$, $\beta$, and $\gamma$.
We define $\nu_{000} = 1$ so that the rescaled amplitude reduces to the unmodified amplitude in the low-energy limit.
The remaining parameters $\nu_{\alpha\beta\gamma}$ are the nuisance parameters that model the mismatch between the amplitude predicted by the  truncated EFT and the true UV model.
If the amplitude involves identical particles, the nuisance parameters must be chosen to respect the crossing symmetry of the amplitude;
for example, if 3 and 4 are identical bosons, the amplitude is invariant under $\hat{s}_{13} \leftrightarrow \hat{s}_{14}$, and we have $\nu_{\alpha\beta\gamma} = \nu_{\alpha\gamma\beta}$.  Another similar complication is that in a global analysis, the same nuisance parameters may affect multiple searches and thus it is important to utilize the same values and sometimes modify which Mandelstam they depend on.   
The function $F(x)$ is a `form factor' that is proportional to $x$ for $|x| \ll 1$, and approaches a constant for $|x| \gg 1$. 
As a baseline, we use the function
\[
F(x) = \begin{cases}  x &  |x| \le 1,
\label{eq:baselineFF}
\\
{\rm{sign}}(x) & |x| \ge 1.
\end{cases}
\]
In the regime $|x_{1,2,3}| \ll 1$ 
we have
\[
\scr{R}_\nu(x_1, x_2, x_3) = 1 + \nu_{100}\; x_1 + \nu_{010}\;  x_2 + \nu_{110}\;  x_1\, x_2 + \cdots\,.
\eql{stu123}
\]
We argued below~\Eq{genM} that in general UV models the Mandelstam descendants coefficients are of order one or less, so the same must be true for the nuisance parameters. 
This is enforced by the  prior likelihood function for the nuisance parameters, as explained below.

The prefactor in \Eq{FFonlyR} is defined so that the modified amplitude has good high energy behavior in all possible scattering channels.
The high energy limit is defined by the regime $|\hat{s}_{12}|, |\hat{s}_{13}|, |\hat{s}_{14}| \sim E^2$ with $E \to \infty$.
(Physically, this corresponds to large angle scattering at high energy.)
In this limit the prefactor scales as $[F(x_\text{max})/x_\text{max}]^{{\mathrm p}/2} \sim E^{-{\mathrm p}}$, and we choose the value of ${\mathrm p}$ so that the modified amplitude is independent of energy in the UV, as we would expect for a scale-invariant UV model.
For example, if the amplitude scales as $\scr{M} \sim \hat{s}_{12}$ at high energies, we choose ${\mathrm p} = 2$, while if the amplitude scales as $\scr{M} \sim \hat{s}_{12} \hat{s}_{13}$, we choose ${\mathrm p} = 4$.

Our form factor modeling of the high energy regime clearly does not offer a complete description of all possible behaviors for the amplitude.
For example, it does not model the possible presence of a resonant peak at high energies. 
The ideal description of the high energy regime would be an agnostic parameterization of the amplitude as an unknown function of the kinematics, provided for instance by a neural network or a Gaussian mixture, subject to basic theoretical requirements, such as perturbative unitarity and crossing symmetry. Exploring this direction is left to future work.
In fact, due to the prefactor, our form factor is arguably not a good approximate description of \emph{any} physical UV model.
Despite this, we will see that the simple 
form
of \Eq{FFonlyR} is already sufficient for many applications.
In particular, we will explain and verify in \S\ref{ss:rFF} that the form factor defined above is sufficient for EFT exclusion analyses, because it is able to mimic the 
predictions of the SM
in the high-energy regime. 
In the presence of a signal, the reduced flexibility of the amplitude modeling could diminish the prospects for a new physics discovery. However, we will see in \S\ref{subDisc} that the potential for discovery of our method is already quite 
good with our baseline
modeling of the high energy regime.

\subsection{The Prior Likelihood}

The $\nu_{\alpha\beta\gamma}$ parameters account for unknown corrections to the EFT predictions.
They are therefore
treated as nuisance parameters in the statistical analysis of the data~\cite{ParticleDataGroup:2024cfk}. In the Bayesian approach, the nuisances are interpreted as statistical variables, with a prior probability distribution, $P_p(\nu)$, that needs to be specified. In the likelihood-based approach, the nuisances are not statistical variables and what needs to be specified is their `prior likelihood', $\mathfrak{L}_p[\nu]$.  
In a strict likelihood interpretation, this prior comes from
an experiment where the nuisance parameters have been measured.  However, for uncertainties of theoretical origin, there is no experiment which measures these parameters.
Therefore the prior likelihood reflects an assumption, not supported by data,
on the most likely value of the true nuisance parameters. In exactly the same way, the Bayesian prior probability distribution $P_p(\nu)$ reflects a belief on $\nu$. As previously explained, we expect the nuisance parameters to be numbers of order one or less and the prior probability distribution or likelihood must reflect this expectation.  

To proceed, we must make this prior quantitative.  Based on experience, it is not difficult to quantify the notion of an `order one' number. An order one number is smaller than a few, in absolute value, with high probability.  Specifically, we impose this expectation by requiring
\[
\label{eq:orderone}
\text{Prob}\big(|\nu| < 3\big) = 0.68\,,
\]
namely that a magnitude bigger than 3 is a 1$\sigma$ deviation, plausible but starting to become unlikely. We have no 
reason to prefer one sign of $\nu$ over another, so
$P_p(\nu)$ should be symmetric in $\nu$. If $P_p(\nu)$ is a Gaussian centered at zero, imposing \Eq{orderone} determines its standard deviation to be equal to $3$. 

Bayesian methods are rarely employed for LHC data analysis, so
in this paper we will adopt the classical likelihood-based methodology where the parameters are not statistical variables and the prior expectation on $\nu$ must be expressed in terms of confidence intervals. The equivalent of \Eq{orderone} is that the interval $[-3,+3]$ is the $1\sigma$ confidence interval for $\nu$, as it emerges from an imaginary measurement. If we measure an estimator that is Gaussian distributed around the true value $\nu$ with standard deviation $\sigma_\nu$ and the result of the measurement is $\bar\nu=0$, the $1\sigma$ confidence interval is the interval  $[-\sigma_\nu,+\sigma_\nu]$, hence we set $\sigma_\nu=3$ to reflect our expectations. The likelihood of this imaginary prior measurement normalized to its maximum (at $\nu=\bar\nu=0$) is given by 
\begin{equation}
\label{Lik}
 \text{Baseline Gaussian prior:} \quad   {-2}\log \frac{\mathfrak{L}_p[\nu]}{\mathfrak{L}_p[\bar\nu]}=\left(\frac{\nu}{\sigma_\nu}\right)^2\,,\quad{\sigma_\nu=3}\,.
\end{equation}
The Gaussian prior likelihood with $\sigma_\nu=3$ defines our baseline for $\mathfrak{L}_p[\nu]$.

Note that with this choice of prior, the most likely value of the nuisance parameters $\nu$ is 0, corresponding to the vanishing of the descendants.
In \S\ref{sec:prior} we will consider an alternative prior that is maximized for $|\nu| \sim 1$ and disfavors $|\nu| \ll 1$.
We will see that the results are very similar for exclusion provided that \Eq{orderone} is imposed.
However, we believe that the Gaussian prior is preferable conceptually because it allows the nuisance parameters to be much smaller than 1, which, as previously explained, is likely to occur in some UV models.

\subsection{Preview of Results and Comparison with Existing Proposals}
\label{sec:comparison}
Let us summarize the important features of our approach and anticipate some of the key results obtained in the rest of the paper:
\begin{itemize}
\item 
The EFT uncertainties are parameterized by modifying the parton-level amplitude with nuisance parameters that reflect the underlying EFT uncertainty in an EFT with cutoff $M$.
\item This modification is
equivalent to a rescaling of the Wilson coefficient, which enables a computationally economical implementation of our method by event reweighting. 
\item
The parameter $M$ that controls the EFT uncertainties has 
a clear physical interpretation as
the EFT cutoff which is of order the mass of new particles in UV realizations.
\item
The prior assumptions limiting the higher order terms are made explicit and have a clear interpretation in terms of UV models.
\item
The constraints converge quickly as more nuisance parameters are added, enabling us to truncate the
polynomial series in Eq.~\ref{eq:FFonlyR} at a small order.
\item
The method gives conservative constraints, namely they are weaker than the constraints on more predictive UV models with new physics at the scale $M$.
\end{itemize}

Let us compare this with Refs.~\cite{Boughezal:2022nof,Allwicher:2022gkm}, which were already mentioned in the introduction.
They considered the constraints from Drell-Yan on dimension-6 SMEFT operators, including dimension 8 operators to estimate the uncertainties.
More specifically, they obtained weakened limits by marginalizing over the coefficients of dimension 8 operators with a flat prior that restricts the couplings to be perturbative up to some scale $M_\text{max}$ (typically in the TeV range). 
In our language, this corresponds roughly to choosing a form factor that is a pure polynomial in the Mandelstam variables, with no softening at high energies.
However, in \S\ref{ss:rFF}, we show that the softening of the amplitude is required to solve the EFT validity issue, in the sense that a polynomial form factor does not eliminate the dependence of the results on the events with energy above the EFT cutoff.
In particular, the right panel of Fig.~\ref{fig:reach_s_baseline_vs_x} shows that with such a form factor the constraints do not become weaker as the EFT cutoff is lowered, in contrast with our method.  Finally, although our analysis does not consider the most general higher order uncertainties (equivalently all higher dimensional operators), our analysis of Mandelstam descendants is easy to implement by using event reweighting and due to the observed rapid convergence with the number of nuisance parameters.

\section{A Concrete Implementation}\label{sec:CI}
This section describes the implementation of our method in a simple toy model, focusing on exclusion. (Discovery will be discussed in \S\ref{subDisc} below.) We consider a model of quarks and leptons interacting via QED and QCD (without weak interactions) and we regard it as our `Standard Model' background. The `new physics' of interest is the EFT operator
\begin{equation}\label{FTEFT}
{\mathcal{L}}_{\textsc{eft}}=-
G(\bar u_L\gamma_\mu u_L)(\bar e_R\gamma^\mu e_R)\,,
\end{equation}
which is one of the interactions that describe weak interactions in the Fermi theory. This simplified setup is chosen because the effective interaction in Eq.~(\ref{FTEFT}) can be UV-completed by the exchange of either a $Z^\prime$ in the $s$-channel or a scalar leptoquark $\phi$ in the $t$-channel. This allows us to explore the dependence of our results on the physics above the EFT cutoff, which we will do in \S\ref{subExc} and \S\ref{subDisc} below. This simple setup will allow us to explain the implementation of our method in a way that can be directly ported to realistic EFT searches at the LHC.

The partonic process $\bar{u} u \to e^+ e^-$ has energy growth $\sim s$, so the modified amplitude \Eq{FFonlyR} is defined with $\mathrm{p} = 2$.
We test our new physics model in the process $pp\to e^+e^-$ at the 13~TeV LHC with an integrated luminosity of 3~ab$^{-1}$. The analysis is performed by counting events in non-overlapping bins and three different binning strategies are considered in order to demonstrate that our method is not dependent on any specific choice of 
kinematic variables:
\vspace{-5pt}
\begin{itemize}
 \setlength\itemsep{-0.3em}
\item{$\boldsymbol{m_{\ell\ell}}${\bf{ binning:}}} the binning is performed on the invariant mass of the dilepton pair, with the bins listed in Table~\ref{tab:xsec_coefficients}.
\item{$\boldsymbol{p_T}${\bf{ binning:}}} the binning is performed on the transverse momentum of the leptons, which are equal for $e^+$ and $e^-$ in our tree-level simulation. The boundaries of the $p_T$ bins are equal to one half of those of the invariant mass bins of Table~\ref{tab:xsec_coefficients}.
\item{$\boldsymbol{3d}$}{\bf{ binning:}} 
the binning is performed on three variables: $p_T$ as defined above, the absolute rapidity $y$ of the dilepton system, the variable $c_*$, defined~\cite{Panico:2021vav} as the cosine of the angle between the $e^-$ momentum and the direction of motion of the lab frame. The $p_T$ bins are those given above, the $y$ bins are given by $[0, 0.1, 0.2, 0.3, 0.4, 0.5,1, 1.5,2.5]$ and the $c_*$ bins are given by $[-1,-0.5, -0.25,0,0.25,0.5,1]$.
This three-dimensional binning strategy exploits the contribution of the EFT operator~(\ref{FTEFT}) to angular distributions, resulting in enhanced sensitivity to $G$.
\end{itemize}
For all binnings, we impose the acceptance cut $p_T>40$~GeV, as well as a cut $|\eta_\ell|<2.5$ on the pseudo-rapidity of the leptons.

\subsection{Statistical Inference}\label{ssec:SI}
Statistical inference is based on the likelihood function for binned data
\begin{equation}\label{eq:lik0}
\mathfrak{L}_{\rm{Po}}[\e;\o]=
\prod\limits_{b\,\in\,{\textrm{bins}}}
{\textrm{Poiss}}\big[\o_b\big|\e_b\big]\,,
\end{equation}
where `Poiss' denotes the Poisson distribution. Here $\o_b$ denotes the observed number of counts in bin $b$, while $\e_b$ is the expected number of events (cross section times integrated luminosity) in that bin. The expected number of events depends on the EFT coupling strength $G$ and on the EFT nuisance parameters $\nu$ defined in Eq.~(\ref{eq:FFonlyR}). It also depends on the cutoff scale $M$, but this dependence is left implicit for brevity and we denote the expected counts as $\e_b=\e_b(G,\nu)$. The complete likelihood is the product of the Poisson term and the prior likelihood for the nuisance parameters
\begin{equation}\label{eq:lik}
\mathfrak{L}[G,\nu;\o]=\mathfrak{L}_{\rm{Po}}[\e_b(G,\nu);\o]\times\mathfrak{L}_p[\nu] \,.
\end{equation}
If not specified otherwise, the prior likelihood $\mathfrak{L}_p[\nu]$ is taken to be the baseline Gaussian given in Eq.~(\ref{Lik}).

We will compute expected exclusion limits for the EFT coupling $G$ (the parameter of interest), for a given value of $M$, under the hypothesis that the observed data are generated by the Standard Model (SM) background. Notice that in the SM 
we have $G = 0$ and 
the dependence on the nuisance parameters drops out. We can thus represent the SM hypothesis as any point on the $G=0$ subspace 
of the $(G,\nu)$ parameter space. It is convenient to think the SM as the point $(0,\bar\nu)$, where $\bar\nu$ denotes the maximum of the prior likelihood for the nuisance parameters ($\bar\nu = 0$ for our baseline prior).
We denote as $\b_b=\e_b(0,\bar\nu)$ the expected number of events in the SM background hypothesis. 

Following the standard LHC statistical practice~\cite{Cowan:2010js,ParticleDataGroup:2024cfk}, we base our inference on the test statistics variable
\begin{equation}\label{eq:tst}
t_G(\o)=2\,\log\frac{\mathfrak{L}[\widehat{G},\widehat{\nu};\o]}{\mathfrak{L}[G,\hathat{\nu}_{G};\o]}\,,
\end{equation}
where $(\widehat{G}, \widehat{\nu})$ denote the values of the parameters that maximize the likelihood function, while $\hathat{\nu}_G$  denotes the values of the nuisance parameters that maximize the likelihood for a specified value of $G$. If the observed data $\o$ were distributed as predicted by the EFT model with an EFT operator coefficient that is equal to $G$, the test statistics $t_G$ follows a $\chi^2$ distribution with one degree of freedom in the large-sample limit, i.e. when the data statistics is large. We can thus compute the $p$-value by comparing $t_G(\o)$ with the cumulative of the $\chi^2$, which in particular for 2$\sigma$ exclusion ($95\%$ confidence level), sets $t_G(\o)>3.84$. The observed data $\o$ are instead assumed to be distributed as predicted by the SM, with 
an expected number of events 
$\b_b$ in each bin. The expected exclusion limit on $G$, $G_{\textrm{exc}}$, is defined as the value of $G$ such that the median of $t_G(\o)$ under the SM hypothesis is equal to $3.84$. 

A commonly-adopted `Asimov' estimate~\cite{Cowan:2010js} for the median is given by the $t_G(\o)$ variable evaluated on observed data that are exactly equal to the expected:
\begin{equation}\label{tstAs}
t_{G}(\b)=2\log\frac{\mathfrak{L}[\widehat{G},\widehat{\nu};\b]}{\mathfrak{L}[G,\hathat{\nu}_G;\b]}
=2\log\frac{\mathfrak{L}_{\textrm{Po}}[\b;\b]}
{\mathfrak{L}_{\textrm{Po}}[\e(G,\hathat{\nu}_G);\b]}
\frac{\mathfrak{L}_{\textrm{prior}}[\bar\nu]}{\mathfrak{L}_{\textrm{prior}}[\hathat{\nu}_G]}
\,.
\end{equation}
Here we simplified the likelihood in the numerator by noticing that the maximum of the Poisson likelihood is when the expected is equal to the observed and the prior likelihood is maximized at $\nu=\bar\nu$. Each factor in the likelihood~(\ref{eq:lik}) is thus independently maximized for $G=\widehat{G}=0$ and $\nu=\widehat\nu=\bar\nu$, when $\o=\b$. Using Eq.~(\ref{tstAs}) we can obtain $G_{\textrm{exc}}$ as the solution to the equation
\begin{equation}\label{limitapprox}
t_{G_{\textrm{exc}}}(\b)=3.84
\,.
\end{equation}
In this paper we will use this to estimate the exclusion reach on $G$.

We will compare our results with a `plain EFT' search,
where the predictions of the EFT operator~(\ref{FTEFT}) are used 
and no nuisance parameters are introduced. The test statistics variable for this `plain EFT' setup is defined as 
\begin{equation}\label{eq:tstPEFT}
t_G^{\textsc{peft}}(\o)=2\,\log\frac{\mathfrak{L}_{\rm{Po}}[\e(\widehat{G});\o]}{\mathfrak{L}_{\rm{Po}}[\e(G);\o]}\,.
\end{equation}
Only the Poisson term is present in the likelihood and the expected counts $\e_b=\e_b(G)$ are the plain EFT predictions. The median of the test statistics in the SM hypothesis is estimated with the Asimov approximation:
\begin{equation}\label{tstPEFTAs}
t_{G}^{\textsc{peft}}(\b)
={2\,\log\frac{\mathfrak{L}_{\textrm{Po}}[\b;\b]}{\mathfrak{L}_{\textrm{Po}}[\e(G);\b]}}
\,.
\end{equation}
In this approximation, 
the exclusion reach will be obtained by comparing with the $2\sigma$ threshold of $3.84$ owing to the $\chi^2$ approximation, as previously explained.  In Appendix \ref{subcorr}, we discuss these issues in further detail.
We also test the Asimov and large-sample approximations, finding that deviations from them on the constraints for $G$ are less than 10\% unless $M< {\rm TeV}$.

\subsection{Event Generation and Reweighting}\label{sec:EGR}

We generate Monte Carlo (MC) events using \texttt{MadGraph5\_aMC@NLO}~\cite{Alwall:2014hca} at tree-level. All the results of the present section employ a single background MC sample of $10^5$ $pp\to e^+e^-$ events. This sample is generated using a UFO model produced with \textsc{FeynRules}~\cite{Alloul:2013bka} that only incudes QCD and QED vertices. The acceptance cuts on the leptons are imposed at the generator level. The total cross section of the generated sample is $\sigma^{\textsc{sm}}=5.28$ pb, corresponding to around $16$ million expected events. In order for the simulation to provide accurate predictions in the entire phase-space, the MC sampling needs to be conditioned by a bias factor that increases the MC statistics in the high-energy tail. A bias proportional to $p_T^4$ is applied by a simple modification of the \texttt{MadGraph} routine \texttt{ptj\_bias}.

The effect of the EFT coupling $G$ and the nuisance parameters is introduced by the technique of event reweighting. The EFT vertex Eq.~(\ref{FTEFT}) is implemented in the UFO model and the automated \texttt{MadGraph} reweighting tool is used to compute the weight 
for each MC event $e$ for a given value of $G$. 
The Feynman amplitude depends linearly on $G$, so the weight is a quadratic polynomial:
\begin{equation}\label{eq:rwg1}
w_e^{\textsc{eft}}(G)=w_e^{\textsc{sm}}\left[
1+{\mathfrak{l}}_e\,\!G+{\mathfrak{q}}_e\,\!G^2
\right],
\end{equation}
where $w_e^{\textsc{sm}}$ are the event weights in the SM, normalized to $\sum_ew_e^{\textsc{sm}}=\sigma^{\textsc{sm}}$, and the linear and quadratic coefficients ${\mathfrak{l}}_e$ and ${\mathfrak{q}}_e$ are computed by reweighting the MC sample with two different nonzero values of $G$. Using Eq.~(\ref{eq:rwg1}), the dependence on $G$ of the cross section in each bin can be determined as the sum of the weights of the events falling in the bin.

\begin{table}[t!]
\centering

\renewcommand{\arraystretch}{0.8}
{
\begin{tabular}{|l|l|l|l|l|l|l|l|l|}
\hline
Bin & $\sigma_b^{\textsc{sm}} $ & $l_{b;000}$  & $l_{b;100}$  & $q_{b;000}^{000}$  & $q_{b;000}^{100}=q_{b;100}^{000}$  & $q_{b;100}^{100}$  \\
$[{\textrm{GeV}}]$ & $[{\textrm{pb}}]$ & $\left[{\text{TeV}^2}\right]$ & $\left[{\text{TeV}^2}\right]$  &  $\left[{\text{TeV}^4}\right]$ &  $\left[{\text{TeV}^4}\right]$  &  $\left[{\text{TeV}^4}\right]$ \\
        \hline
        $[80, 150)$ & $3.9$ & $-6.5 \ 10^{-2}$ & $-8.9 \ 10^{-4}$ & $7.1 \ 10^{-3}$ & $1.1 \ 10^{-4}$ & $2.0 \ 10^{-6}$ \\
        $[150, 200)$ & $0.76$ & $-1.9 \ 10^{-1}$ & $-5.6 \ 10^{-3}$ & $4.9 \ 10^{-2}$ & $1.5 \ 10^{-3}$ & $4.8 \ 10^{-5}$ \\
        $[200, 300)$ & $0.44$ & $-3.6 \ 10^{-1}$ & $-2.2 \ 10^{-2}$ & $0.18$ & $1.2 \ 10^{-2}$ & $8.1 \ 10^{-4}$ \\
        $[300, 400)$ & $0.12$ & $-6.4 \ 10^{-1}$ & $-7.7 \ 10^{-2}$ & $0.63$ & $7.9 \ 10^{-2}$ & $1.0 \ 10^{-2}$ \\
        $[400, 500)$ & $3.9 \ 10^{-2}$ & $-1.2$ & $-0.24$ & $1.9$ & $0.40$ & $8.2 \ 10^{-2}$ \\
        $[500, 600)$ & $1.7 \ 10^{-2}$ & $-1.8$ & $-0.53$ & $4.4$ & $1.3$ & $0.40$ \\
        $[600, 800)$ & $1.2 \ 10^{-2}$ & $-3.1$ & $-1.5$ & $12$ & $6.0$ & $3.1$ \\
        $[800, 1000)$ & $3.7 \ 10^{-3}$ & $-5.2$ & $-4.1$ & $34$ & $27$ & $22$ \\
        $[1000, 1500)$ & $2.1 \ 10^{-3}$ & $-7.7$ & $-7.7$ & $63$ & $63$ & $63$ \\
        $[1500, 2000)$ & $3.1 \ 10^{-4}$ & $-7.2$ & $-7.2$ & $59$ & $59$ & $59$ \\
        $[2000, 3000)$ & $7.9 \ 10^{-5}$ & $-7.2$ & $-7.2$ & $59$ & $59$ & $59$ \\
        $[3000, 13000)$ & $7.0 \ 10^{-6}$ & $-8.0$ & $-8.0$ & $66$ & $66$ & $66$ \\
        \hline
\end{tabular}
}
\caption{The coefficients that parametrize the dependence on the $\nu_{100}$ nuisance in Eq.~(\ref{eq:exp}) for $m_{\ell\ell}$ binning. The cutoff is set to $M=1$~TeV.}
\label{tab:xsec_coefficients}
\end{table}

The EFT nuisance parameters $\nu$ can now be introduced as a phase-space dependent rescaling of the EFT operator coefficient, as in Eq.~(\ref{eq:Rdefn}). 
The dependence of the weights on $\nu$ is then obtained from the linear and quadratic weights coefficients ${\mathfrak{l}}_e$ and ${\mathfrak{q}}_e$ of Eq.~(\ref{eq:rwg1}), given the event-by-event knowledge of the truth-level partonic kinematical variables.
In this case, only a single scattering channel contributes, and we use the  notation
\[
x = x_1 = \frac{\hat{s}}{M^2},
\quad
y = x_2 = \frac{\hat{t}}{M^2},
\quad
z = x_3 = \frac{\hat{u}}{M^2}.
\]
Specifically, the weights are
\begin{equation}\label{eq:rwg2}
w_e^{\textsc{eft}}(G,\nu)=w_e^{\textsc{sm}}\left[
1+{\mathfrak{l}}_e\,\!\scr{R}_\nu(x_e, y_e, z_e)\,G+{\mathfrak{q}}_e\,\!(\scr{R}_\nu(x_e, y_e, z_e))^2G^2
\right],
\end{equation}
where $e$ labels the event.
The cross section in each bin, as a function of both the parameter of interest $G$ and the nuisance parameters $\nu$, is the sum of the weights of the events falling in the bin. Its determination only requires the values of ${\mathfrak{l}}_e$ and ${\mathfrak{q}}_e$ defined in Eq.~(\ref{eq:rwg1}), which can be obtained with standard tools as previously described. Relative to a standard EFT analysis where only ${\mathfrak{l}}_e$ and ${\mathfrak{q}}_e$ are needed, the inclusion of nuisance parameters in the prediction of the cross sections comes without any 
significant additional computational cost or complication. 

The implementation of our strategy is even simpler if the amplitude rescaling function $\scr{R}_\nu$ is linear in the nuisance parameters, as in Eq.~(\ref{eq:FFonlyR}). In this case, the weights~(\ref{eq:rwg2}) depend polynomially both on $G$ and on $\nu$, as does the cross section in each bin. We can thus express analytically the expected number of events in each bin, $\e_b(G,\nu)$, in the compact form
\begin{equation}\label{eq:exp}
\e_b(G,\nu)=
\e_b^{\textsc{sm}}\Bigg[
1+
G\sum\limits_{\alpha,\beta,\gamma}
(l_b)_{\alpha\beta\gamma} 
\nu_{\alpha\beta\gamma}+
G^2 \!\!
\sum\limits_{\substack{\alpha,\beta,\gamma \\ 
\alpha'\!, \beta'\!, \gamma'}}
(q_b)_{\alpha\beta\gamma}^{\alpha'\! \beta'\! \gamma'}
\nu_{\alpha\beta\gamma}
\nu_{\alpha'\!\beta'\!\gamma'}
\Bigg]\,,
\end{equation}
where $\e_b^{\textsc{sm}}$ is the expected number of events in the background-only model and the polynomial coefficients $l_b$ and $q_b$ are obtained by summing weights in each bin:
\begin{equation}
\begin{split}
\label{eq:coeff}
&(l_b)_{\alpha\beta\gamma}=\frac1{\sigma^{\textsc{sm}}_b}
\sum\limits_{e\,\in\, b}
w_e^{\textsc{sm}}{\mathfrak{l}}_e
\left( \frac{F({x_{\text{max},e}})}{x_{\text{max},e}}\right)^{\frac{\textrm{p}}{2}}
F({x_e})^\alpha
F({y_e})^\beta
F({z_e})^\gamma\,,
\\[8pt]
&
\!\!\!\!
(q_b)_{\alpha\beta\gamma}^{\alpha'\!\beta'\!\gamma'}
= \frac1{\sigma^{\textsc{sm}}_b}
\sum\limits_{e\,\in\, b} % \sum\limits_{e\in b}
w_e^{\textsc{sm}}{\mathfrak{q}}_e
\left( \frac{F({x_{\text{max},e}})}{x_{\text{max},e}}\right)^{{\textrm{p}}}
F(x_e)^{\alpha + \alpha'}
F(y_e)^{\beta + \beta'}
F(z_e)^{\gamma + \gamma'}
\,.
\end{split}
\end{equation}
Here $x_{\text{max}_e} = \text{max}\{ x_e, y_e, z_e\}$, and $\mathrm p$ is defined below \Eq{FFonlyR}.

Table~\ref{tab:xsec_coefficients} lists for illustration the predictions for the SM cross section $\sigma_b^{\textsc{sm}}$ and the $l$ and $q$ coefficients in the case of the $m_{\ell\ell}$ binning,
for a cutoff scale $M = 1$~TeV. The table gives the coefficients required to include the effects of $\nu_{100}$, the first nuisance parameter in the $s$-channel. The different components of the $l$ and $q$ tensors are equal in the bins where $m_{\ell\ell}>M=1$~TeV because 
$F(x) = 1$ for $x > 1$ (see Eq.~(\ref{eq:baselineFF})). 

We remind the reader 
that in our notation $\nu_{000}$ is not a nuisance parameter; it is fixed to $\nu_{000} = 1$, since it multiplies the leading effect of the EFT coupling $G$ at low energies. With this convention, the $000$ term of the first sum in Eq.~(\ref{eq:exp}) and the $000,000$ term of the second sum are independent of the nuisance parameters.  They parameterize the nuisance-independent contributions to the events in the bin proportional to $G$ and $G^2$, respectively. Notice also that Eq.~(\ref{eq:exp}) depends implicitly on the  EFT cutoff parameter $M$ through the form factor $F$, which is a function of $x = x_1 = \hat{s}/M^2$, $y = x_2 = \hat{t}/M^2$, and $z = x_3 = \hat{u}/M^2$ (see Eqs.~(\ref{eq:Rdefn}) and~(\ref{eq:FFonlyR})). This dependence cannot be parameterized analytically by a simple function, so the parametrization~(\ref{eq:exp}) is for a fixed $M$, and the $l$ and $q$ coefficients must be recomputed for each value of $M$ used 
in the analysis. Similarly, the coefficients depend implicitly on the functional form of the form factor. The baseline form factor~(\ref{eq:baselineFF}) is employed in all the results of the present section if not specified otherwise.

The expected number of events in each bin serves as the input for the calculation of the likelihood function~(\ref{eq:lik}), which in turn is used to obtain the expected limit using Eq.~(\ref{limitapprox}). We determine $G_{\textrm{exc}}$ defined in Eq.~(\ref{limitapprox}) numerically by employing standard functions in the \texttt{SciPy.optimize} package~\cite{2020SciPy-NMeth}. Specifically, we use \texttt{minimize} (the `Nalder--Mead' method) to compute the maximum over the nuisances of the log-likelihood ratio, as a function of $G$, and \texttt{brentq} to solve Eq.~(\ref{limitapprox}) for $G_{\textrm{exc}}$. This was found to be sufficient to produce all the results of this paper on a commercial laptop computer.

It is worth commenting on the scalability of our method to real EFT searches based on LHC data. Our starting point are the weight coefficients ${\mathfrak{l}}_e$ and ${\mathfrak{q}}_e$ that parametrize the dependence on the EFT coupling $G$ as in Eq.~(\ref{eq:rwg1}). These can be obtained with established automated tools that are already employed in ATLAS and CMS analyses~\cite{ATLAS:2023eld,ATLAS:2024hmk,ATLAS:2024hac,ATLAS:2025yww,CMS:2014fjm,CMS:2017ret,CMS:2017egm,CMS:2019efc,CMS:2019nep,CMS:2019ppl,CMS:2020lrr,Mestdach:2021nki,Cruz:2023bmn,CMS:2023xyc,Basnet:2023xml,CMS:2024jvu,CMS:2025xkn,CMS:2025ugn} in order to incorporate the dependence on the EFT Wilson coefficients in the predictions. For instance, the \texttt{MadGraph} suite interfaced with EFT UFO models such as SMEFT~\cite{Degrande:2020evl} or SMEFTsim~\cite{Brivio:2017btx} enables the generation of truth-level events in the SM and the determination of the EFT polynomial coefficients for an arbitrary dimension-6 SMEFT 
operator up to NLO in QCD. Passing the generated truth-level sample through the simulation of the detector response produces fully realistic detector-level events that retain all the information on the truth-level coefficients and on the truth-level kinematics.
Using this information, the dependence on the nuisance parameters is included by reprocessing the data with Eq.~(\ref{eq:rwg2}) or by directly computing $\e_b(G,\nu)$ in the truth-level bins using Eq.~(\ref{eq:exp}), if the dependence on the nuisance is polynomial. Nuisance parameters accounting for other uncertainties, such as PDF uncertainties, scale variation or experimental errors, can be incorporated straightforwardly. We believe that the inclusion of the EFT nuisance parameters in the theoretical predictions does not pose any significant additional challenges in comparison with standard EFT searches.

The same conclusion holds in the case of an analysis constraining several EFT operators simultaneously. Our simulation strategy is straightforward to generalize to this case, as discussed in Appendix~\ref{app:MO}. An independent set of nuisance parameters must be introduced for each operator, but the results we present in the next section suggest that only one or two nuisance
parameters will be needed for each operator. Hence, the total number of parameters should not exceed what can be handled by the powerful minimization tools employed by the experimental collaborations.

\subsection{Nuisance Parameters}\label{subsec:nup}

The 
main elements of our methodology are the 
specification 
of nuisance parameters to account for the EFT truncation uncertainties, 
a prior likelihood for the nuisance parameters,
and the presence of form factors that regulate the behavior of the amplitude above the cutoff scale. The role played in the analysis by each of these three elements is described in this and in the following two sections in turn.

\begin{figure}[t]
    \centering
    \includegraphics[width=1\linewidth]{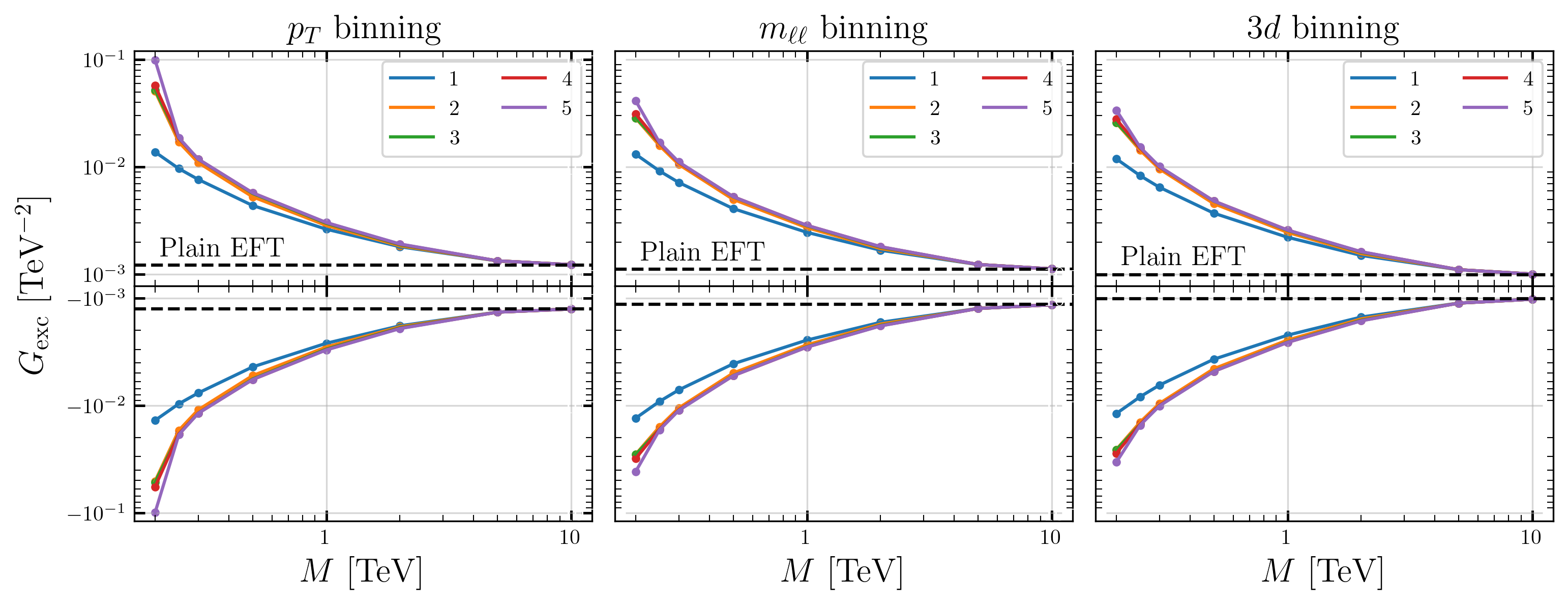}
    \caption{The 95$\%$CL exclusion reach on the EFT operator Wilson coefficient for a number of $s$-channel nuisance parameters varying from 1 to 5, as a function of the cutoff $M$. The horizontal dashed `plain EFT' lines are the reach of a regular analysis that does not include nuisance parameters. The $p_T$, $m_{\ell\ell}$ and $3d$ binnings are considered in the three panels.}
    \label{fig:3binnings}
\end{figure}

Fig.~\ref{fig:3binnings} displays the $2\sigma$ expected exclusion reach for positive and negative $G$ as a function of the cutoff scale $M$ in the three binning schemes previously described ($m_{\ell\ell}$, $p_T$, and $3d$ binning). The reach is similar for the different binnings, with the bounds for the $3d$ binning strategy only slightly stronger (by around $20\%$), as expected. A notable feature of these results is that the reach converges rapidly as the number of included nuisance parameters is increased, especially for the the largest values of $M$ which are of the most physical interest for EFT models. These results include only `$s$-channel' nuisance parameters, namely $\nu_{\alpha00}$ with $\alpha$ ranging from 1 to 5. The inclusion of additional nuisance parameters in the $t$- and $u$-channels have a small effect on these results, as we will discuss later.

To understand the role of the nuisance parameters, we consider the partonic process $u_L{\bar{u}}_R \to e_R \bar{e}_L$, which contributes
to our dilepton final state.
The contributions from the SM and the EFT operator~(\ref{FTEFT}) to the amplitude are given by
\begin{equation}\label{eq:amps}
{\mathcal{M}}_{\textrm{SM}}= {-\tfrac{4}{3}} e^2 \sin^2(\theta/2)\,,
\qquad
{\mathcal{M}}_{\textrm{EFT}}=2G{\hat{s}}\sin^2(\theta/2)\,,
\end{equation}
where $e$ is the electric charge, $\theta$ is the scattering angle and $\hat{s}=E^2$ is the partonic center of mass energy squared. The EFT operator contribution to the amplitude grows quadratically with the energy $E=\sqrt{\hat{s}}$. Our methodology replaces the constant Wilson coefficient $G$ with a kinematics-dependent function as in Eq.~(\ref{eq:Rdefn}). The partonic scattering amplitude that we are effectively employing for our predictions thus reads
\[
%\!\!\!\!\!\!\!\!
{\mathcal{M}} &= {\mathcal{M}}_{\textrm{SM}}
+2G{\hat{s}}\sin^2(\theta/2) 
\frac{F(\hat{s}/M^2)}{\hat{s}/M^2} 
\Big[
    1+\nu_{100}F(\hat{s}/M^2)+
    \nu_{200}F(\hat{s}/M^2)^2
    + \cdots \Big].
\label{eq:expf}
\] 
The square of this amplitude, integrated over the phase space, produces the theoretical predictions for the cross sections in each bin, as a function of the Wilson coefficient $G$, the cutoff $M$ and the nuisance parameters.

The term proportional to $G$ in Eq.~(\ref{eq:expf}) produces departures from the background-only predictions and the minimization over the nuisance parameters tries to make its effect as small as possible in order to mimic the background-only data. For events with 
$\hat{s} \ll M^2$, the prefactor of the square brackets is just
${\mathcal{M}}_{\textrm{EFT}}$.
For this kinematics, the `$1$' term in the square bracket gives the pure EFT prediction and the other terms are small corrections.
Therefore, for events of this kind, the nuisance parameters do not cancel the departures from the background.
Adding more nuisance parameters makes even smaller corrections to the amplitude in this regime, so the predictions of the model are robustly close to those of the plain EFT at low energies.

For events with $\hat{s} \gsim M^2$ the EFT loses predictivity.
Therefore, if these high-energy events are compatible with the SM predictions, this should not constrain $G$.
The 
form factor is essential to achieve the cancellation of the new physics contribution in this regime.
With our baseline choice for the function $F(x)$ (see \Eq{baselineFF}),
the prefactor of the square bracket is a constant equal to the plain EFT amplitude at $\hat{s} = M^2$ and therefore does not have any high energy growth for $\hat{s} > M^2$.
Therefore, the new physics contribution can be completely canceled for all $\hat{s} > M^2$ by choosing order one nuisance parameters to cancel the `1' in the square brackets.  
It is therefore not surprising that the results converge quickly as we add more nuisance parameters, as illustrated in Fig.~\ref{fig:3binnings}.

Not using a form factor is equivalent to setting $F(x) = x$ in \Eq{FFonlyR}.
In this case, for events with $\hat{s} \gsim M^2$, the high-energy growth in the coefficient of the square brackets is not canceled;
in fact, the modified amplitude contains higher powers of $\hat{s}$, making the high-energy behavior worse.
In this case, order one nuisance parameters are not sufficient to cancel the EFT contribution to the amplitude in the regime $\hat{s} \gsim M^2$.
We study this setup in \S\ref{ss:rFF} and we find that in fact the results do not converge as we add more nuisance parameters (see Fig.~\ref{fig:reach_s_baseline_vs_x}, right panel).

The parameter $M$ is the mass scale that separates the low-energy and high-energy regimes.
If $M$ is sufficiently large, all of the data is in the low-energy regime and the nuisance parameters have little effect on the predictions.
Maximizing the likelihood in this regime simply sets the nuisance parameters to the maximum of their prior likelihood ($\nu = \bar\nu)$.
The statistical inference based on the limit $M \to \infty$ is therefore 
identical to the plain EFT search.
This behavior is clearly seen in Fig.~\ref{fig:3binnings}, where the results of our method converge to the plain EFT results for large $M$.
The difference between the two results gives a quantitative estimate of the uncertainty due to higher order EFT corrections in this regime.

For lower values of $M$, more and more of the data is in the high-energy regime, where the nuisance parameters can adjust to make the predictions agree with the SM.
Therefore, these searches effectively limit the data used to constrain the model to events with $\hat{s} \ll M^2$, giving a conservative bound based only on the predictions of the EFT in the kinematic regime where these are reliable.  We also see that one or two nuisance parameters are sufficient for a satisfactory estimate of the reach over the whole range of $M$. The reason is that the first nuisance parameters---specifically, $\nu_{100}$ or $\nu_{010}$ or $\nu_{001}$---give the dominant contribution to the predictions smearing, of order $F(E^2/M^2)=E^2/M^2$ in the bins with a characteristic energy $E<M$. For sufficiently low values of $M$, the EFT loses all predictivity and we expect the limit on $G$ to diverge. Correspondingly, the reach at the lowest $M$ point in Figure~\ref{fig:3binnings} is more than 10 times larger than at large $M$ and it would further increase if the plot was extended to even lower $M$. However, we are not really interested in the quantitative outcome of our analysis when $M$ is so low that all data are beyond the regime of validity of the EFT.

The nuisance parameters effectively `smear out' the EFT prediction in accordance with the theoretical uncertainties of the EFT framework. 
It is therefore less predictive than the `plain EFT' and therefore should give weaker constraints. 
This is particularly clear in the Asimov approximation Eq.~(\ref{tstAs}), where the observed are equal to the background-only expectations. The maximization performed on the 
likelihood in the denominator 
determines the value of the nuisance parameters $\hathat{\nu}_G$ that best fit the data and since the data are background-only this tends to eliminate from the predictions any departures from the background-only model (the SM) that are induced by the presence of the EFT operator. Therefore, the predictions for $\hathat{\nu}_G$ are closer to the SM predictions than the plain EFT. The plain EFT test statistics in Eq.~(\ref{tstPEFTAs}) does not have adjustable parameters in the denominator likelihood and therefore it is larger than our test statistics in Eq.~(\ref{tstAs}) for any given $G$. Hence, our limit is more conservative than the plain EFT.

\begin{figure}
    \centering
    \includegraphics[width=0.65\linewidth]{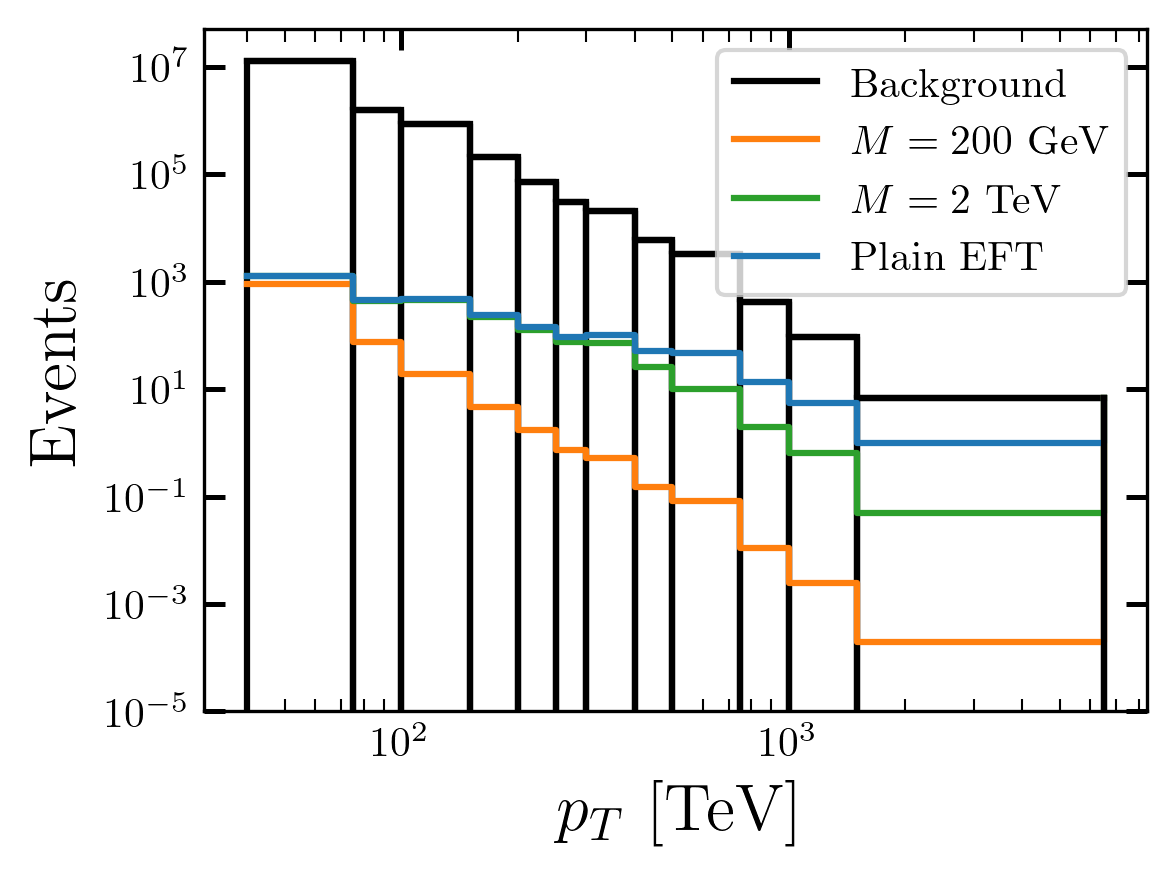}
    \caption{The number of events in excess to the background, in the $p_T$ bins, for the best-fit value of the nuisance parameters when $M=200$~GeV and $M=1$~TeV, and in the plain EFT analysis. The total number of background events is also reported.
    }
    \label{fig:Background_vs_EFT_at_fixed_g}
\end{figure}

This is illustrated in Fig.~\ref{fig:Background_vs_EFT_at_fixed_g}, which shows number of events in each $p_T$ bin for the background model, as well as the excess events predicted by different EFT models with $G=-1.2\times 10^{-3}\ \text{TeV}^{-2}$. (This value of $G$ corresponds to $2\sigma$ exclusion for the plain EFT search.) We see that including the nuisance parameters does not dramatically affect the predictions for bins with $p_T < M$, but it significantly reduces the number of predicted events for bins with $p_T \gsim M$, in accordance with the discussion above.

\begin{figure}
    \centering
    \includegraphics[width=\linewidth]{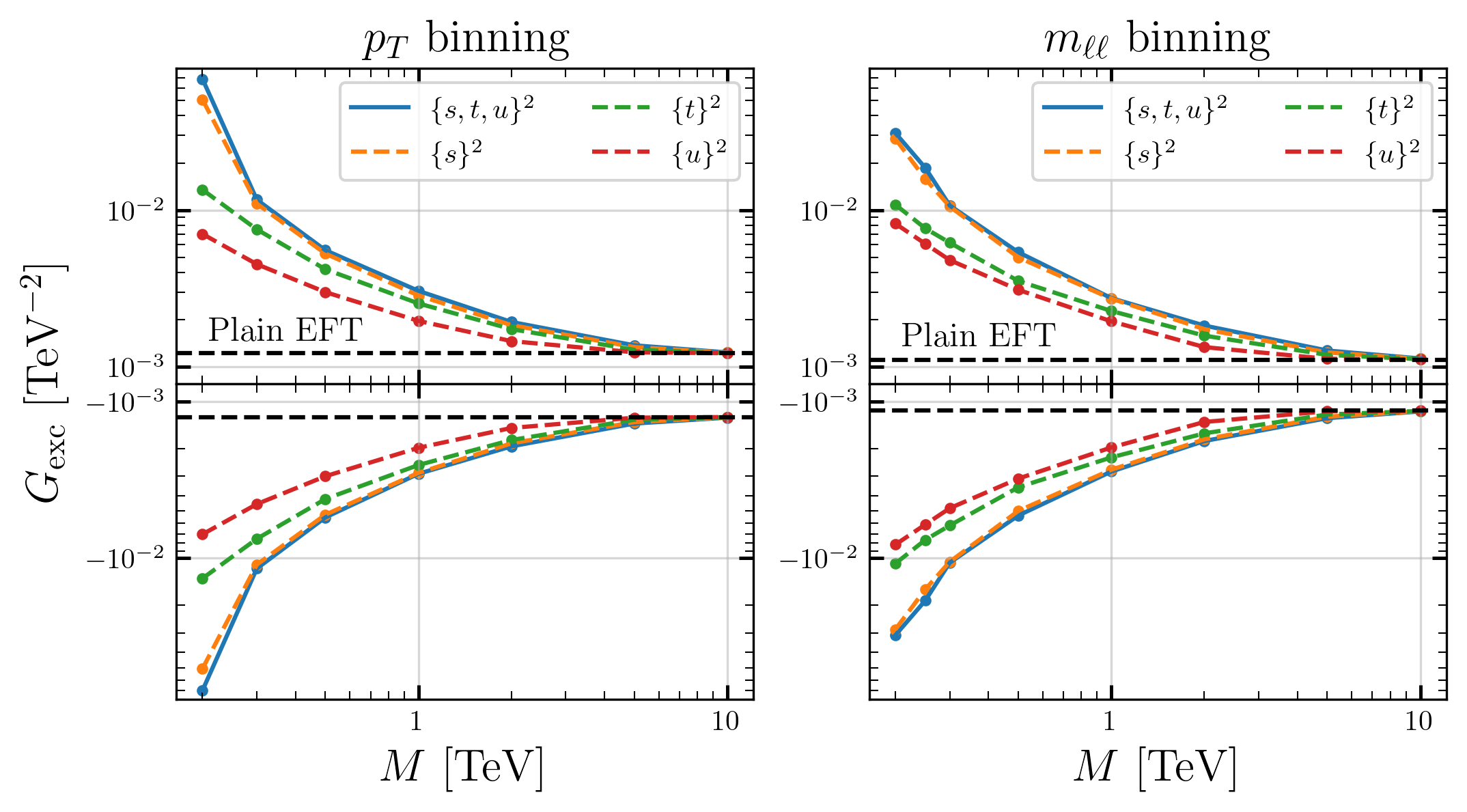}
    \caption{95$\%$~CL reach, as a function of the cutoff $M$, in 4 different nuisance parameters configurations that correspond to the most general degree-2 polynomial in the three Mandelstam channels ($\{s,t,u\}^2$) and to degree-2 polynomials ($\{s\}^2$, $\{t\}^2$, $\{u\}^2$) in the individual channels. 
    }
    \label{fig:s_t_u_nuisances}
\end{figure}

The results discussed so far make use only of $s$-channel nuisance parameters.
Fig.~\ref{fig:s_t_u_nuisances} shows how the reach changes if we include $t$- and $u$-channel nuisance parameters. The dashed lines show the reach if we include 2 nuisance parameters in a single channel; of these, the $s$-channel limits are the most conservative. The solid line shows the reach including nuisance parameters up to quadratic order in all three channels (9 total nuisance parameters). These bounds are practically indistinguishable from the case where only $s$-channel nuisance parameters are included. This means that the $s$-channel nuisance parameters are more effective than the others in canceling the effect of the EFT operator at high energies; equivalently, they imply larger uncertainties in the EFT prediction at high energies. This is easily understood from the elementary fact that $\hat{s} \ge |\hat{t}|, |\hat{u}|$, which makes the $s$-channel nuisance parameters more important event by event. This suggests that $t$- and $u$-channel nuisance parameters can be omitted in analyses aimed at exclusion. However, $t$- and $u$-channel nuisance parameters should be considered for analyses aimed at discovery, since they are expected to improve the modeling of the data in cases with new physics in the $t$- or $u$-channel.

\subsection{Dependence on the Prior}\label{sec:prior}
\begin{figure}
    \centering
    \includegraphics[width=\linewidth]{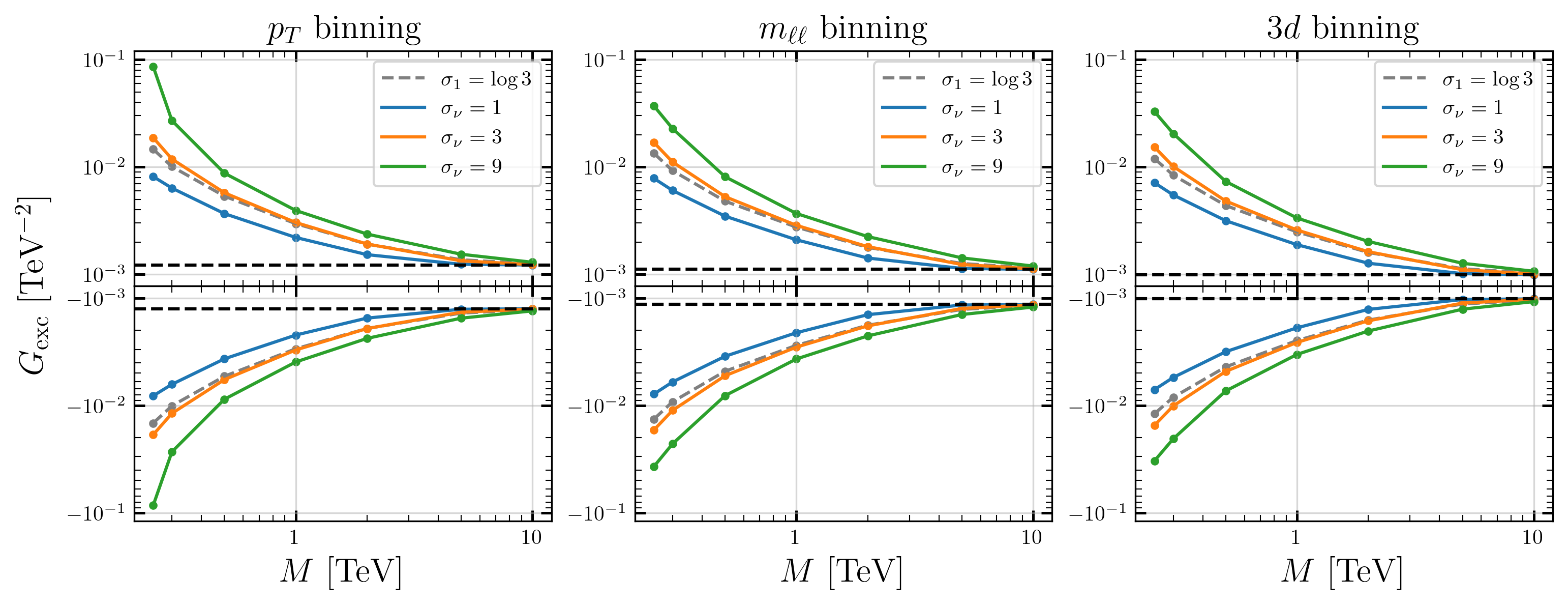}
    \caption{95$\%$~CL reach, as a function of the cutoff $M$---using $5$ $s$-channel nuisance parameters---for different choices of the nuisance parameters prior. Gaussian priors  with $\sigma_\nu=1$ and $9$ are compared with the baseline $\sigma_\nu=3$. The log-normal prior with standard deviation $\sigma_{\rm{l}}=\log3$ is also considered.}
\label{fig:reach_s_diff_prior_norm}
\end{figure}

The considerations of the previous section rely implicitly on the assumption that the nuisance parameters cannot be much larger than one. Otherwise, their contribution to the cross section 
can be arbitrarily large, independently of the  
energy and the cutoff scale $M$. 
A setup where the nuisance parameters are completely 
unconstrained 
effectively corresponds to assigning order-one theoretical uncertainties in all data 
bins, independent of $M$, making it impossible to extract any information on $G$ from the 
data. 

In our setup, nuisance parameters are effectively bounded by the presence of the prior likelihood in Eq.~(\ref{Lik}). This suppresses the total likelihood~(\ref{eq:lik}) when the nuisance parameters are comparable or larger than $\sigma_\nu$, making the likelihood maximization in the test statistics~(\ref{tstAs}) behave as follows. On the one hand, the nuisance need to reproduce the data as closely as possible in order to maximize the Poisson component on the likelihood. If this can be achieved in a configuration where $\nu$ is much smaller or at most comparable with $\sigma_\nu$, the prior likelihood has a minor impact and the nuisance parameters can freely adjust to the data. Otherwise, the prior likelihood prevents the nuisance parameters to attain the large value that would be required to fit the data. Both behaviors can be observed by looking at the green line on the right panel of Fig.~\ref{fig:Background_vs_EFT_at_fixed_g}, obtained with $M=2$~TeV. A good cancellation of the EFT effects---reproducing the background-only data---is achieved in those bins with a characteristic energy scale of the order of the cutoff scale, while no cancellation is possible at much lower energy bins. This is because in the latter bins the effect of the nuisance parameters is suppressed by $E^2/M^2$, a very large value of the parameters would be needed for the cancellation and this is forbidden by the presence of the prior likelihood.

The fact that the constraints depend on a choice of prior likelihood for the higher order EFT corrections is unavoidable, as we have already stressed several times.
Any experimental protocol that extracts a bound on EFT parameters is therefore  
making some assumption about the size of higher order corrections.
The goal of our method is to make these prior assumptions explicit and their meaning transparent.
With our method, it  
is straightforward to compare the results obtained with different choices of prior to aid in the interpretation of the results.

As an example of this, we study the effect of varying the allowed size of the nuisance parameters $\nu$.
Our baseline prior 
is a Gaussian with width $\sigma_\nu = 3$ (see Eq.~\ref{Lik}).
The dependence on the width of the distribution is illustrated in 
Fig.~\ref{fig:reach_s_diff_prior_norm}, which shows the expected exclusion reach for $\sigma_\nu = 1, 3, 9$.
(This figure also shows the results for a log-normal prior, which will be described below.)
As expected, the bounds are weaker for large values of $\sigma_\nu$.
To interpret these results, recall that the baseline choice $\sigma_\nu=3$ emerges from an assessment of what it means for a nuisance parameter to be an `order-one number.' 
Specifically, we assume that the probability for an order one number to exceed 3 is $32\%$, corresponding to a $1\sigma$ fluctuation.
The choice $\sigma_\nu = 9$ corresponds to assuming that 9 is a typical order one number, while for $\sigma_\nu = 1$ the probability of finding 3 or larger has probability $0.003$.
It is clear that $\sigma_\nu = 1$ is too restrictive and we would argue that $\sigma_\nu = 9$ is too permissive.
A reasonable procedure for experimental searches would be to report the results for $\sigma_\nu = 3$ and $9$.

While the results must and do depend on the quantitative notion of `order-one' that is assumed, 
they should
be insensitive to the functional form of the prior likelihood.
To check this, we introduce the log-normal prior
\begin{equation}
\label{Lik1}
\text{Log Normal Prior:}  \quad  {-2}\log \frac{\mathfrak{L}_p[\nu]}{\mathfrak{L}_p[\bar\nu]}=\left(\frac{\log|\nu|}{\sigma_{\rm{l}}}\right)^2\,,
    \quad{\rm{with}}\quad{\sigma_{\rm{l}}=\log3}\,.
\end{equation}
The maximum of the likelihood is now at $\nu=\bar\nu=1$ and the probability for small values of $\nu$ is suppressed.
By setting $\sigma_{\rm{l}}=\log3$, $z$ ranges from $-3$ to $+3$ with 
approximately $68\%$ probability, in accordance with \Eq{orderone}.%
\footnote{The agreement is not exact, because in the log-normal prior the probability is $68\%$ to be $\nu\in[-3,-1/3]\cup [1/3,3]$. However we ignore this small difference for simplicity.}
Fig.~\ref{fig:reach_s_diff_prior_norm} shows that the log-normal prior with $\sigma_{\rm{l}}=\log3$ produces essentially identical results as the Gaussian prior with $\sigma_\nu=3$ for exclusion.
The interpretation of this is that for exclusion, the most important thing is the upper bound on the nuisance parameters, as explained above.
The impact of the shape of the prior is less clear for discovery, as we will discuss in \S\ref{subDisc} below.

\subsection{The Shape of the Form Factor}
\label{ss:rFF}
In \S\ref{sec:howto} and \S\ref{ss:formfactors} we argued that a form factor is required to take into account the fact that the EFT loses predictivity for $E \gsim M$, where $M$ is the EFT cutoff.
In this section we discuss the dependence on the choice of form factor $F(x)$ employed in our method.

\begin{figure}
    \centering
    \includegraphics[width=\linewidth]{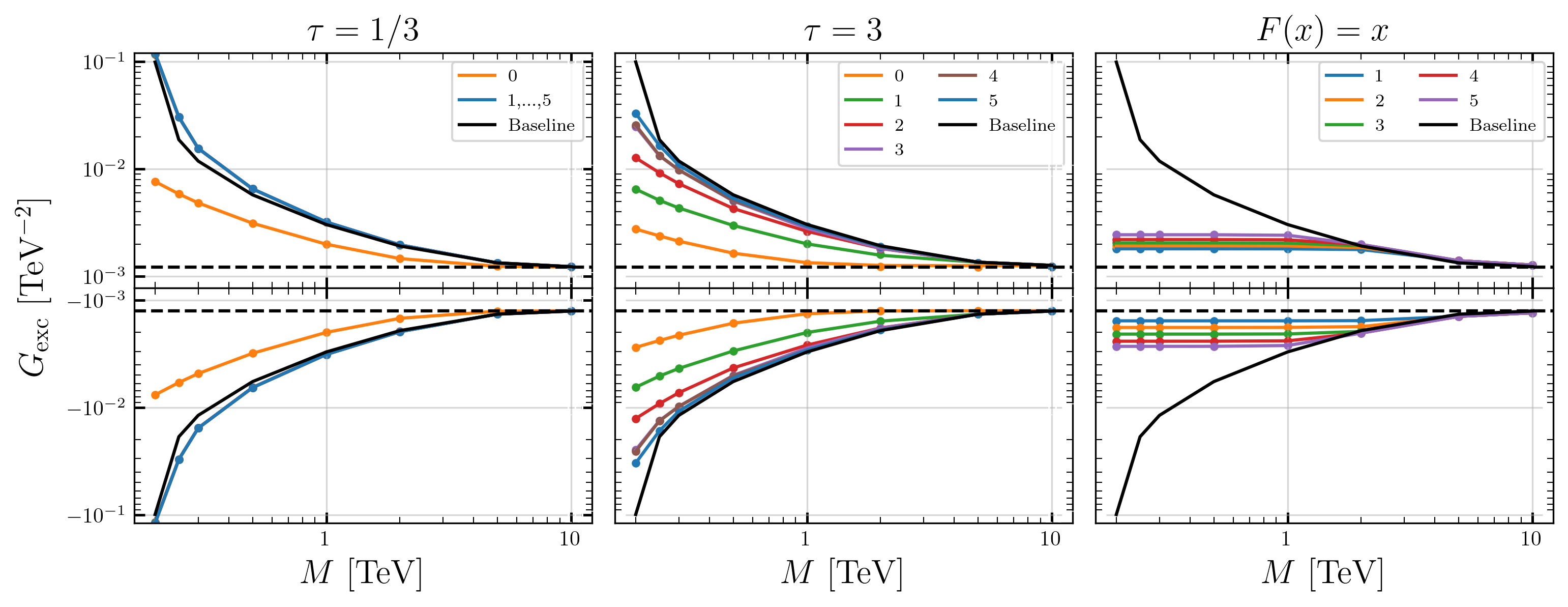}
    \caption{95$\%$~CL reach of the $p_T$-binning analysis, for three different choices---see the main text---of $F(x)$. The dashed colored lines display the reach for a variable number of $s$-channel nuisance parameters. For $\tau=1/3$, the reach is the same for any number of parameters ranging from 1 to 5 and is labeled as `$1,...,5$'. The lines marked as `0' do not employ nuisance parameters. The reach of the plain EFT analysis and of the baseline setup with $5$ nuisance parameters are reported for reference in each plot as black dashed and  continuous lines, respectively.}
    \label{fig:reach_s_baseline_vs_x}
\end{figure}

We start by considering the case where higher derivative corrections are included without a form factor, {\it i.e.}~the choice $F(x) = x$ in \Eq{FFonlyR}.
We note that this is equivalent to adding a finite number of higher derivative operators.
We will show that there are serious problems with this approach and we expect similar problems in any approach where additional EFT operators are included to assess the EFT uncertainties.
The key point is that in this case, the amplitude has polynomial growth at large energies, so that the model violates unitarity.
Of more direct relevance to us, this setup produces exclusion limits that are clearly unphysical, as can be seen in the right panel of Fig.~\ref{fig:reach_s_baseline_vs_x}.
In particular, the bounds become independent of $M$ for sufficiently small $M$, while it is clear that they should be becoming weaker as a function of $M$, since the model is supposed to be less predictive.

\begin{figure}
    \centering
    \includegraphics[width=\linewidth]{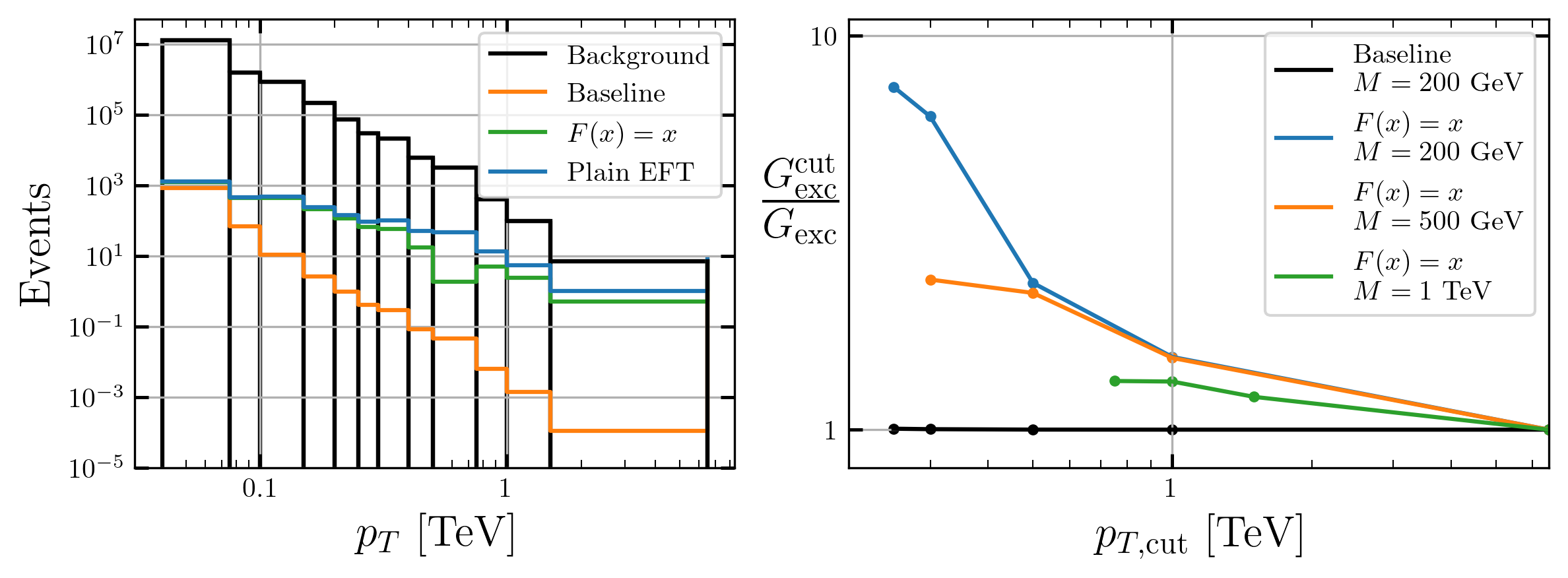}
\caption{{\bf{Left:}} Same as the left panel of Fig.~\ref{fig:Background_vs_EFT_at_fixed_g} for $M=200$~GeV, but now including an analysis without form factors, labeled as `$F(x)=x$'. {\bf{Right:}} The variation of the upper limit on $G$ when 
a `clipping' cut $p_T < p_{T,\text{cut}}$ 
is imposed to exclude data with a characteristic energy that is above the cutoff. }
\label{fig:Background_vs_EFT_at_fixed_g_clipping}
\end{figure}

We claim that the reason for this behavior is that the nuisance parameters are not able to cancel the EFT contribution to the amplitude for $E \gsim M$.
This is illustrated in Fig.~\ref{fig:Background_vs_EFT_at_fixed_g_clipping}.
The left panel of the figure is the same as Fig.~\ref{fig:Background_vs_EFT_at_fixed_g} for $M=200$~GeV, but now includes the case where no form factor is used.
From this we see that the predictions for the high energy bins are close to those of the plain EFT, showing that the nuisance parameters are not able to cancel the unphysical behavior for $E \gsim M$.
Furthermore, these bins strongly contribute to the test statistic and hence to the exclusion reach.
This is shown in the right panel of Fig.~\ref{fig:Background_vs_EFT_at_fixed_g_clipping}, which shows the effect on the results of imposing a `clipping' cut $p_T < p_{T,\text{cut}}$ on the data for $M = 200,\,500,\,1000$~GeV.
We see that the cut strongly affects the bound for the lower values of $M$.
For comparison, we show that the effects of the clipping cut on our baseline analysis with $M = 200$~GeV is completely negligible, as we would expect from the fact that the nuisance parameters do effectively cancel the new physics effects for events in the high energy bins in this case.

We now generalize this to study the functional form of the form factor.
We will study a 1-parameter family of form factors given by
\[
F_\tau(x) = \begin{cases} x &  |x| < \tau,
\\
\tau \, {\rm{sign}}(x) 
& |x| \ge \tau.
\end{cases}
\]
These form factors generalize our baseline by having transition from $F(x) = x$ to $F(x) = \text{constant}$ at $x = \tau$ rather than $x = 1$.
The reach for $\tau=\frac 13$ and $\tau=3$ is compared with the reach for our baseline choice $\tau = 1$ in the two leftmost panels of Fig.~\ref{fig:reach_s_baseline_vs_x}.
The reach for a large number of nuisance parameters is essentially identical.
The only significant difference with the baseline is that the convergence as a function of the number of nuisance parameters is slower for $\tau = 3$ and faster for $\tau = \frac 13$ compared with the baseline setup (as seen by comparing to Fig.~\ref{fig:3binnings}).

We conclude that while our methodology requires a form factor, the results are robust under significant variations of its functional form.
This is in contrast to the more traditional use of form factors to cut off the unphysical high energy behavior of models, where the results do depend on the functional form chosen for the form factor or equivalently on the unitarization prescription~\cite{Garcia-Garcia:2019oig}.
This is illustrated in Fig.~\ref{fig:reach_s_baseline_vs_x} by the orange lines (marked `0') in the two leftmost panels.
These are obtained by setting to zero all nuisance parameters, which retains an overall form factor for the amplitude, given by $F_\tau(x)/x = 1/x$ for $x > \tau$.
The limit obtained in this way behaves qualitatively like our proposal, namely the exclusion becomes weaker for smaller values of $M$, as it should.
However, the value of the limit is sensitive to the choice of form factor, as can be seen by comparing the first and second panels.
Once we include sufficiently many nuisance parameters, the different form factors all give essentially identical results.

\section{Exclusion, Discovery, and Comparison with UV Models}
\label{sec:Res}
In this section we present some results of our method for both exclusion and discovery that serve to test the method.
We also compare our method to `data clipping' and searches for specific UV models.

An important claim that we are making is that the results of our method with an EFT cutoff $M$ can be robustly interpreted as conservative 
constraints  
on UV models with new physics at the scale $M$.
In order to test this, we will compare our results with two simple UV models whose leading interaction at low energies is given by the EFT interaction Eq.~(\ref{FTEFT}).
The first model is a new heavy vector $Z^\prime$ coupled to left-handed up quarks and right-handed electrons:
\begin{equation}
    \label{eq:LUV_Zp}
    \mathcal{L}_{\textsc{int}}^{(s)} = g Z^\prime_\mu\left({\bar{u}}_L\gamma^\mu u_L + \bar{e}_R\gamma^\mu e_R\right)\quad\Rightarrow\quad G=\frac{g^2}{M_{Z^\prime}^2}\,.
\end{equation}
In the process $u \bar{u} \to \bar{e} e$ this operator comes from integrating out the $Z'$ in the $s$-channel, so we refer to this as an `$s$-channel' UV-completion.
The second model is a color-triplet scalar leptoquark $\phi$ of charge $\frac 53$, which produces the same EFT interaction, but from a $t$-channel exchange:
\begin{equation}
\label{eq:LUV_LQ}\mathcal{L}_{\textsc{int}}^{(t)} = y\big(\phi\, \bar{u}_L e_R + \phi^\dagger\, \bar{e}_R u_L \big)\quad\Rightarrow\quad G=\frac{y^2}{2M_\phi^2}\,.
\end{equation}
Note that we do not need to preserve the SU$(2)_L\times$U$(1)_Y$ gauge symmetry because only the QED and QCD interactions are present in our simplified setup. 

We will compare the results of our method to these models, where we identify $M$ with either $M_{Z'}$ or $M_\phi$.
We will first discuss exclusion, then discovery.
For exclusion, we also compare our method to `data clipping' in \S\ref{subExcDC}.

\subsection{Setting Limits}\label{subExc}

As explained in \S\ref{ssec:SI}, we estimate the expected $95\%$~CL exclusion limit on the EFT Wilson coefficient $G$ by a hypothesis test based on the profile likelihood test statistics by employing the large-sample ($\chi^2$) and Asimov approximations. We saw in \S\ref{subsec:nup} that the exclusion reach converges quickly for an increasing number of nuisance parameters. 
We found that two parameters of the  $s$-channel type are sufficient for convergent results, so this is what we will use in the following.
We restrict our attention to the $3d$ binning analysis setup that produces slightly stronger constraints. The results are shown on the third panel of Fig.~\ref{fig:3binnings}.

We now compare the EFT exclusion limits with the limits that can be set directly on the $Z^\prime$ and leptoquark  UV-completions, which are obtained as follows. We fix the mass of the heavy particle ($M_{Z^\prime}$ or $M_\phi$) and we compute the theoretical predictions as a function of the heavy particle coupling ($g$ or $y$). This defines a one-parameter statistical model with Poisson likelihood as in Eq.~(\ref{eq:lik0}) and no prior, since there are no nuisance parameters. A maximum-likelihood test statistics is defined similarly to Eq.~(\ref{eq:tstPEFT}), with maximization over the parameter of interest ($g$ or $y$) in the numerator and no maximization in the denominator. The observed counts are set to those expected in the SM background hypothesis owing to the Asimov approximation, therefore the maximum of the likelihood in the numerator is at $g=\widehat{g}=0$ or $y=\widehat{y}=0$. This produces a simple analytic formula for the test statistics, which we set to the threshold of $3.84$ in order to determine the expected $2\sigma$ exclusion limits on the parameters: $g_{\textrm{exc}}$ or $y_{\textrm{exc}}$. The procedure is repeated for different values of the heavy particle mass obtaining mass-dependent coupling exclusions. 
Finally, we use the relations in Eqs.~(\ref{eq:LUV_Zp}) and~(\ref{eq:LUV_LQ}) to convert the limit on the couplings into a limit on the value of the EFT Wilson coefficient $G$ that is generated by the UV model under consideration after integrating out the resonance. We denote these limits as $G_{\textrm{exc}}^{(s)}$ or $G_{\textrm{exc}}^{(t)}$ when they are obtained with the $Z^\prime$ $s$-channel UV completion or with the leptoquark $t$-channel UV completion, respectively.
\begin{figure}
    \centering
    \includegraphics[width=0.65\linewidth]{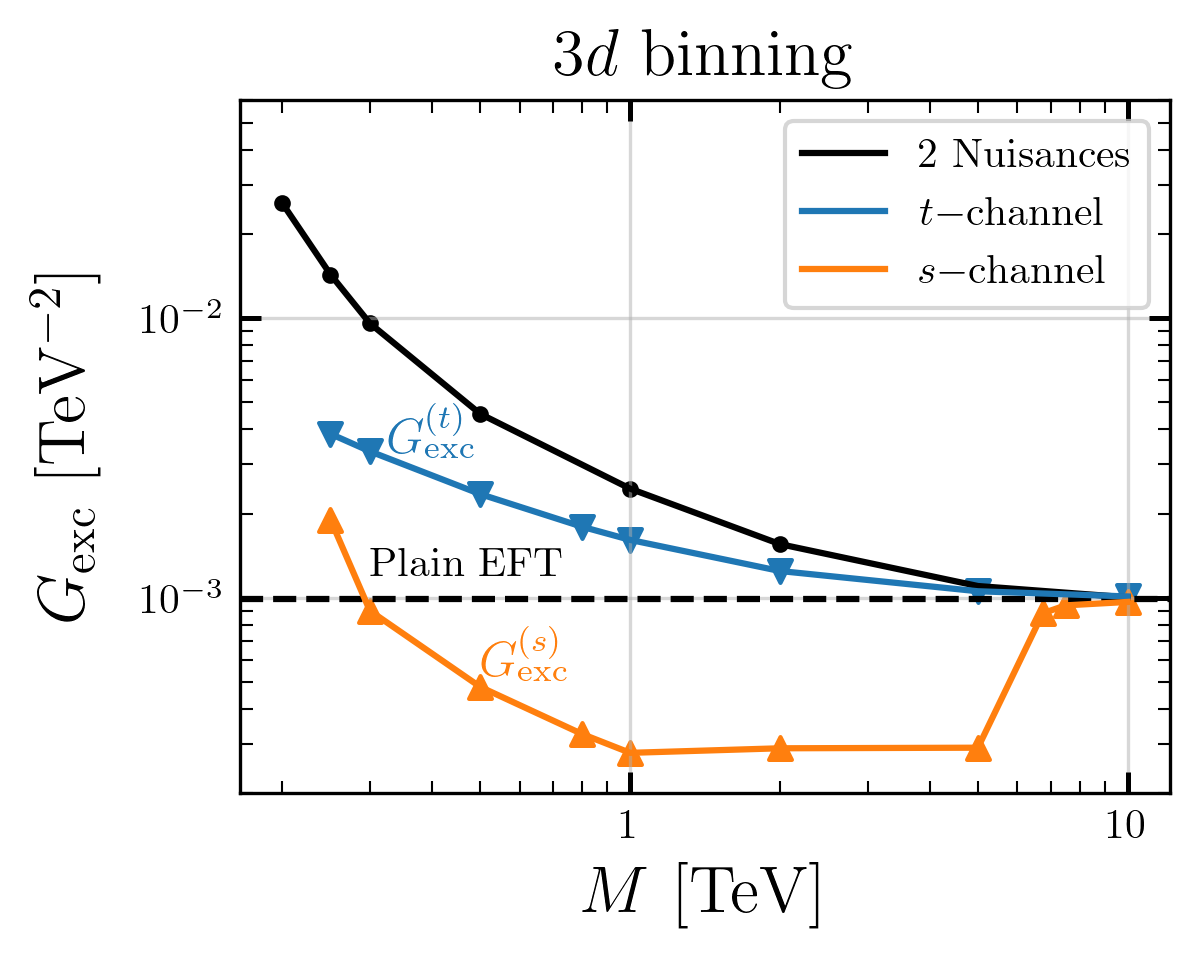}
    \caption{The 95$\%$~CL expected upper bound on $G$ of the $3d$-binning analysis in the plain EFT (dashed black) and with our method using 2 nuisance parameters (solid black) are compared with the results of dedicated searches for the underlying $s$- and $t$-channel UV models.}
    \label{fig:uv_reach}
\end{figure}
In Fig.~\ref{fig:uv_reach} we report a comparison between the exclusion limit obtained with our methodology (black line), with a standard EFT analysis (dashed line), and with the limits obtained in the UV models reported as a  orange (blue) line for the $s$-channel ($t$-channel) UV completion. The horizontal axis is the cutoff $M$ of the EFT, which is identified with the heavy particle mass ($M_{Z^\prime}$ or $M_\phi$) in the explicit UV models. Only the positive upper bound on $G$ is displayed in the figure because the UV models under consideration only populate the $G>0$ region of the EFT parameter space. 

Let us first restrict our attention to the comparison between the EFT limits obtained with our methodology (black solid line) and the UV model limits $G_{\textrm{exc}}^{(s)}$ and $G_{\textrm{exc}}^{(t)}$. Our limit is weaker than the UV-model limits and this is consistent: the model-independent EFT analysis is asking a less specific question to the data than the analysis that is specifically targeted to assess the UV-model viability. Therefore, it can extract less information on the specific model and weaker limits. 
Although the EFT limits are weaker, they are useful because they hold
for a wide class of models, including the many models for which no dedicated search is performed. 

\begin{figure}
    \centering
    \includegraphics[width=\linewidth]{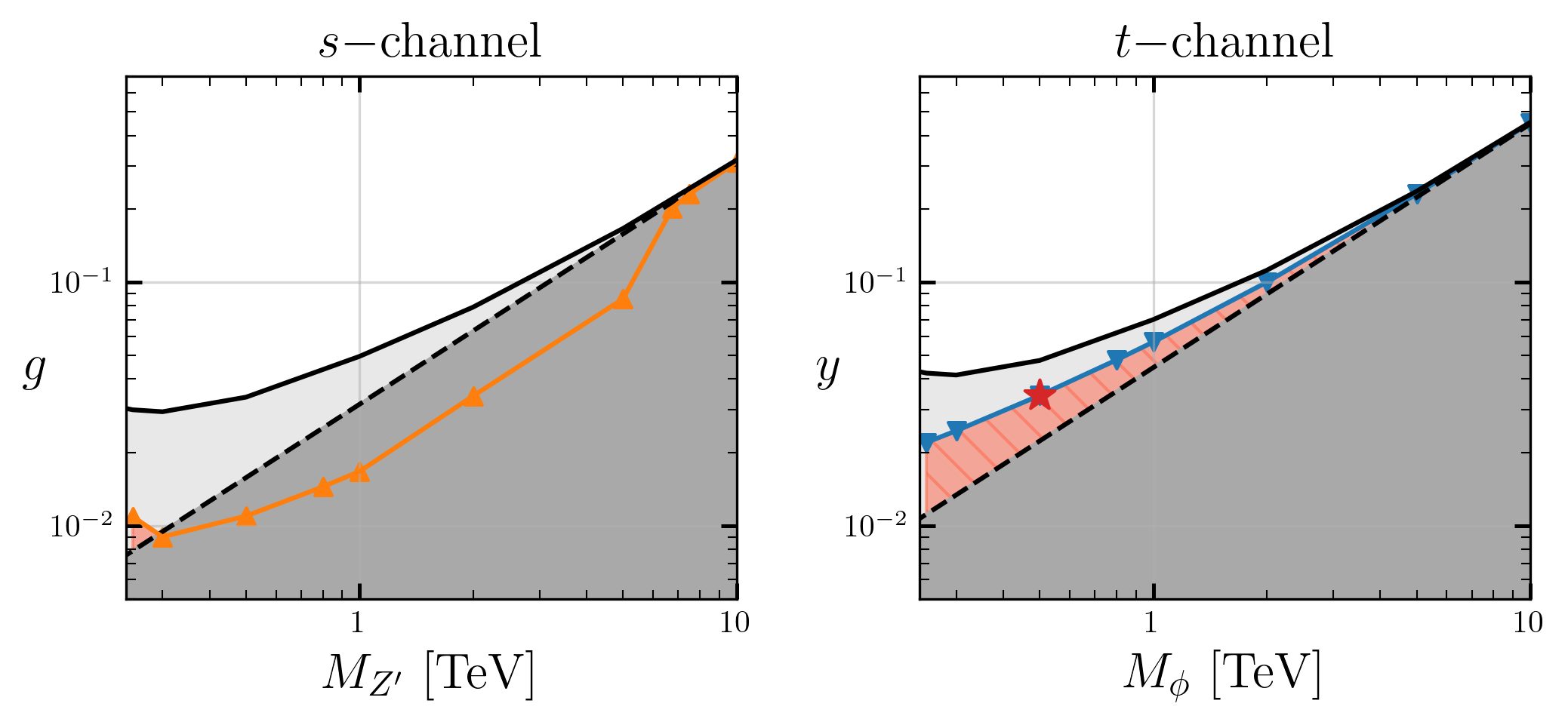}
    \caption{The regions in the $s$-channel (left panel) and in the $t$-channel (right panel) model parameter space that are allowed by the EFT search performed with our method (light gray), with the plain EFT approach (dark gray) and by the UV model search (orange and blue). The red star model will be referenced in Fig.~\ref{fig:uv_reach_fixed_g}. }
    \label{fig:uv_reach_M_vs_g}
\end{figure}

We stress again that it 
is very important that the EFT exclusion limits are \emph{weaker} and not stronger than the UV-model limits.
The reason is that  
this prevents us from drawing incorrect 
conclusions on the viability of UV models. 
To illustrate this point, in Fig.~\ref{fig:uv_reach_M_vs_g} we project the EFT limits in the 2-dimensional parameter space $(M_{Z^\prime},g)$ of the $s$-channel UV model (left panel) and  in the $(M_{\phi},y)$ parameter space of the $t$-channel model (right panel).  In the plots, we compare them with the limits obtained directly within the UV models. The EFT analysis performed with our methodology excludes a portion of the parameter space, leaving an allowed region for the parameters shaded in light gray in the plots. These allowed regions will be all that we know about  the two UV models until dedicated searches are performed. With dedicated searches, we would obtain more information on the model and restrict its allowed parameter space even further as shown in the plots (blue and yellow lines).

This is not the case for 
the plain EFT limit, reported in Fig.~\ref{fig:uv_reach} as a dashed black straight line. The limit is \emph{stronger} than the $t$-channel UV model limit for all values of $M$,
and stronger than the limit for the $s$-channel model for low values of $M$.
Both of these UV models clearly belong to the class of `reasonable' UV models that we would hope to probe with an EFT analysis, so this shows that we cannot interpret the plain EFT analysis as a direct constraint on UV models.
Doing so can lead us to `exclude' new physics that may in fact be present in the data.
This is illustrated in Fig.~\ref{fig:uv_reach_M_vs_g}, which shows the region of the UV model parameter space that is excluded by a direct search, compared to the plain EFT search and our method.
The red shaded hatched region is excluded by the plain EFT search, but allowed by the UV model search.

\begin{figure}
    \centering
    \includegraphics[width=\linewidth]{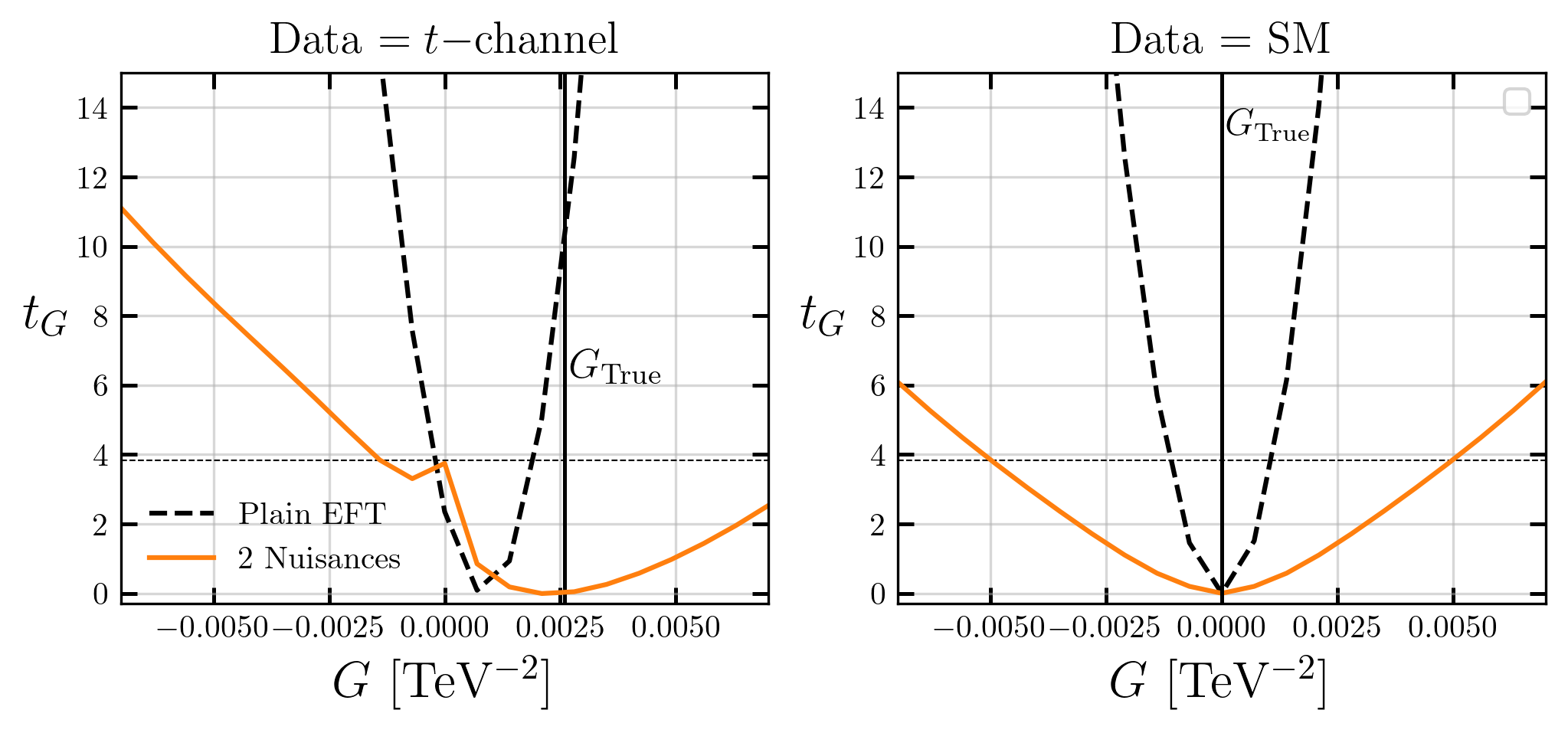}
    \caption{{\bf Left:} The test statistics for exclusion $t_G$ when the observed data are equal to the prediction of the $t$-channel model for $M_\phi=500$~GeV and $y=0.036$. The corresponding value of $G=y^2/(2M_\phi^2)$ is shown as a vertical line. The threshold for $2\sigma$ exclusion, of $3.84$, is also shown. {\bf Right:} The same plot when the observed data is equal to the prediction of the SM background.
    \label{fig:uv_reach_fixed_g}}
\end{figure}

A major risk associated with 
the plain EFT analysis is that
the analysis excludes 
a value of $G$ that is actually realized in Nature. 
To verify that this can indeed happen, 
we must depart from the statistical setup we have been using for exclusion, which assumes 
that the data follow the background-only distribution. We consider the $t$-channel model and set its parameters at the point marked by a red star on the right panel of Fig.~\ref{fig:uv_reach_M_vs_g}, namely $M_\phi=500$~GeV and $y=0.036$. If the data are distributed as predicted by the model, a good (Asimov) estimate of the typical value of the test statistics observed in experiments is obtained by assuming that the result of the experiment is equal to the model's prediction. The test statistics of the plain EFT analysis in Eq.~(\ref{eq:tstPEFT}), computed with the observed counts equal to the counts predicted by the $t$-channel model, is displayed on the left panel Fig.~\ref{fig:uv_reach_fixed_g} as a function of the value of $G$ that is being tested for exclusion. The test statistics of our analysis, defined as in Eq.~(\ref{eq:tst}) is also plotted for comparison. 
On the right panel of the figure, the same plot is shown but instead taking the observed counts to be equal to the background-only expected, the same setup where the exclusion limits were derived. The interception of the test statistics with the threshold of $3.84$ identifies the $2\sigma$ exclusion reach on $G$ and it coincides with the reach reported in Fig.~\ref{fig:uv_reach} at $M=500$~GeV for the plain EFT and for our analysis.

The right panel of Fig.~\ref{fig:uv_reach_fixed_g} represents the result of the experimental analysis when 
the data is described by the SM. 
In this case, 
the test 
statistic is minimized at $G = 0$ in both the plain EFT and our method, as it should be.
In the right panel,
new physics is present in the data and the low energy EFT coupling has a true value of $G=y^2/(2M_\phi^2)=2.6 \times 10^{3}$~TeV$^{-2}$, 
compatible with 
the parameters chosen for the UV model. 
In our method,  
the minimum of the test statistic is close to the true 
value of $G$, which is marked with a dashed vertical line in the plot. 
On the other hand, the 
minimum of the test statistic for the plain EFT search  
is below the true value and 
grows quickly with $G$. 
At the true value of $G$ the test statistic for the plain EFT search 
is well above the threshold for exclusion.
We conclude that the plain EFT search incorrectly excludes the value of $G$ that is 
present in the data.

These results illustrate concretely the generic risk associated with an analysis that ignores the limited energy range of validity of the EFT. As discussed in the Introduction, the risk stems from the growth with energy of the EFT amplitude that produces large departures from the background-only predictions in the high-energy bins. This growth is an artifact of the unjustified usage of the EFT description beyond the cutoff and 
does not match the behavior of the amplitude in physical models.  
It is true that the plain EFT search gives correct results if $M$ is sufficiently high, but it provides no way of determining how large a value of $M$ is required, or quantitatively estimating the error due to the neglect of higher order terms.
Our method does estimate these uncertainties, and shows that the plain EFT search is valid only for very large values of $M$.

It is not surprising 
that the plain EFT analysis is more prone to fail for $t$-channel rather than $s$-channel UV models. In the $t$-channel model, the true amplitude growth saturates at an energy of the order of the cutoff, while 
the $s$-channel model has a prominent 
peak from the on-shell resonance production.
\begin{figure}
    \centering
    \includegraphics[width=0.5\linewidth]{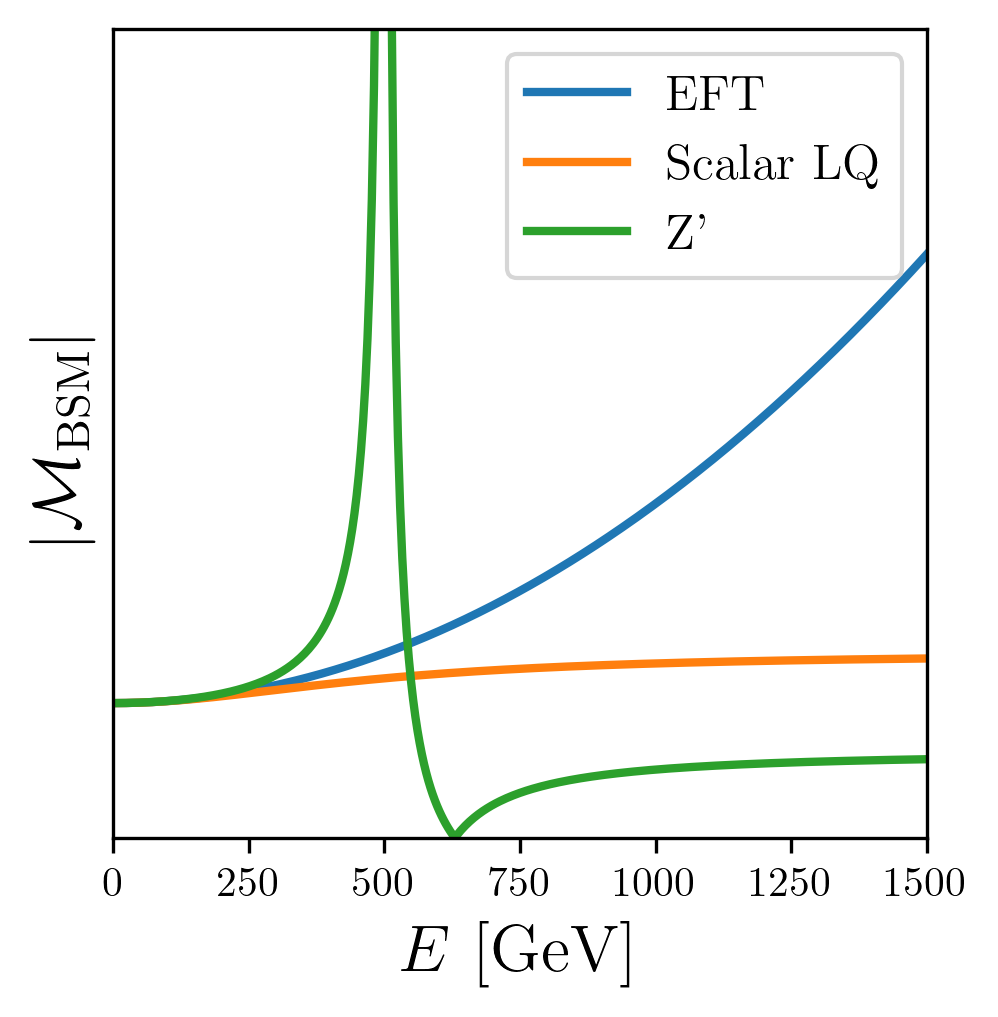}
    \caption{The new physics contribution to the amplitude in the EFT (blue line), in the $s$-channel (orange) and $t$-channel (green) UV completions, where $G=2.6\cdot10^{3}$~TeV$^{-2}$ and in the UV models, the new particle mass is 500 GeV.
}
    \label{fig:amplitude}
\end{figure}
This is shown in Fig.~\ref{fig:amplitude} for two representative values of the resonance mass $M_{Z^\prime}=M_{\phi}=500$~GeV, with the couplings $g$ or $y$ chosen by Eqs.~(\ref{eq:LUV_Zp}) and~(\ref{eq:LUV_LQ}) to match the EFT predictions with the same value of $G=2.6\times 10^{3}$~TeV$^{-2}$. Below the cutoff, both models display the amplitude behavior proportional to $G E^2$ that is predicted by the EFT. At the cutoff, the $s$-channel amplitude is larger than the EFT prediction, so  
the limit in the true model is typically stronger than the EFT limit. This is indeed what we observe in Fig.~\ref{fig:uv_reach} almost over the entire $M$ range. In the $t$-channel model the amplitude is instead smaller than the EFT prediction at high energies, making the true theory predictions less discrepant from the background-only prediction than the plain EFT. 
That is, the incorrect plain EFT predictions give 
erroneously stronger sensitivity.

\subsection{Comparison with Data Clipping}\label{subExcDC}
As we already discussed, the `data clipping' methodology for EFT searches suffers from a number of shortcomings that are addressed in our proposal.
One issue is that measuring the hard scale of the partonic scattering process experimentally may be difficult or impossible (in the case of invisible particles in the final state).
This problem is absent in our toy analysis, since we are working with a parton-level analysis for a leptonic final state.
Even assuming that an accurate measurement of the partonic energy scale is possible, 
the data clipping method still suffers 
from the inherently qualitative connection between the clipping scale $M_{\rm{clip}}$, above which the data are eliminated, and the physical EFT cutoff $M$. In data clipping, $M_{\rm{clip}}$ defines a sharp threshold that separates the region of validity of the EFT from the region where the EFT is not applicable. 
The data above $M_{\rm{clip}}$ are excluded from the analysis because it is not accurately described by the EFT, while for the data below $M_{\rm{clip}}$, the EFT prediction is included with no theoretical uncertainties.
This is not consistent with the identification $M_{\rm{clip}}=M$ because the EFT predictions do suffer from truncation uncertainties and are not fully trustworthy even if the energy is below the cutoff. 
Our method takes the EFT cutoff $M$ as an input parameter and gives a fully quantitative statistical estimate of the impact of the associated EFT uncertainties on the EFT searches for events below the cutoff, whereas data clipping has to exclude a large part of this 
regime to ensure accurate predictions.

\begin{figure}
    \centering
    \includegraphics[width=0.65\linewidth]{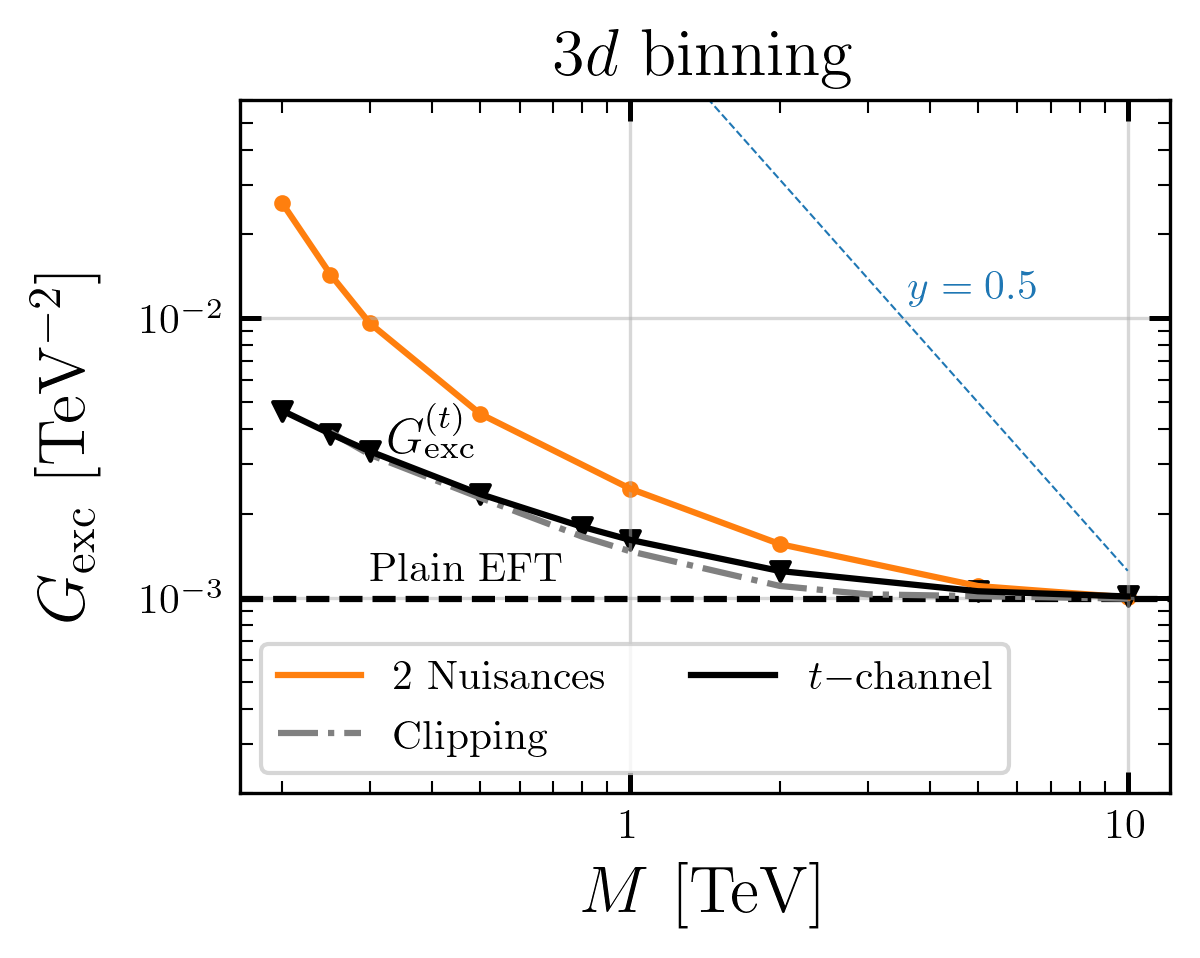}
    \caption{The 95$\%$~CL expected upper bound on $G$ of the $3d$-binning analysis in the plain EFT (dashed black), with our method using 2 nuisance parameters (solid orange)  and with data clipping (solid gray) are compared with the results of a dedicated search for the underlying $t$-channel UV model (solid black). For the data clipping line, $M$ is identified with the clipping scale.  The contour where the UV model coupling $y$ equals $0.5$ is also displayed, showing that the exclusion reach is well within the perturbative regime and far from the perturbative unitarity limit $y\lesssim4\pi$.}
    \label{fig:uv_reach_with_clipping}
\end{figure}

Because of this,  
we expect 
at best qualitative agreement 
between the limits set using our method and those based on data clipping. This is illustrated in Fig.~\ref{fig:uv_reach_with_clipping} for the  study at hand. The analysis with data clipping (gray line) is a regular EFT search, without nuisance parameters, restricted to the bins where the dilepton invariant mass $m_{\ell\ell}$ is smaller than the clipping scale $M_{\rm{clip}}$. The clipping scale is identified with the cutoff $M$. The data clipping exclusion limit is stronger than the reach of our method, as expected. It is weaker than the plain EFT and therefore in less strong tension with the model-specific limit $G_{\rm(exc)}$. However, the inconsistency with the model-specific search is still present in the mass range from around 700~GeV to 4~TeV, where the reach of the clipped EFT analysis is slightly stronger than $G_{\rm(exc)}^{(t)}$.
One could try to turn around and use 
the results of UV model searches 
to identify the correct $M$ to use for a given $M_{\rm{clip}}$.  
However, this identification will depend on  
the UV model(s) chosen, as well as the details of the clipping analysis ({\it e.g.}~cuts and luminosity), 
again highlighting the 
indirect and inherently qualitative  
relationship between 
the clipping scale and the scale of new physics.

\subsection{Discovery}\label{subDisc}
So far we have focused on exclusion, but our statistical model can be straightforwardly deployed in a discovery analysis; an analysis aimed at quantifying the significance of a possible data discrepancy from the background-only SM predictions. It is particularly important to study discoveries within our framework in order to test the possible 
objection that, since our method is more conservative than the plain EFT for exclusions, it must be less effective for discovery. However, this is too na\"ive, because our approach provides a more faithful modeling of the theoretical predictions than the plain EFT. 
This improved modeling 
results in weaker exclusions because it 
correctly 
accounts for the theoretical uncertainties, but can also provide a more accurate fit to a  
discrepancy between the data and the background-only model, which can help discovery.  
However, this improved fit comes 
at the cost of additional parameters and is dependent on our choice of the priors, so it is not clear {\it a priori} how it compares to the plain EFT.
In addition, the comparison between the two methods depends on the UV dynamics that truly underlies the distribution of the data, in particular on the mass of the new particles.   
These issues illustrate that there are additional challenging questions that must be addressed to assess the effectiveness of our method for discovery. 
In this section we investigate 
some of these issues quantitatively, 
using the $t$-channel UV model introduced at the beginning of the section.

A discovery analysis is a test of hypothesis aimed at excluding the SM at a level of confidence that is typically set to $5\sigma$. The effectiveness of the analysis can be quantified by the power of the test when the data are truly distributed according to some given model with fully specified parameters, or quantified by the discovery reach on one of the parameters of the model. We assume the $t$-channel UV models to be the true data distribution, for fixed values of the new particle mass $M_\phi$, and we trade the remaining parameter $y$ with the EFT operator coefficient $G$ using Eq.~(\ref{eq:LUV_LQ}). The discovery reach $G_{\textrm{dis}}$ is the minimal value of $G$ that enables, if truly realized in Nature, the $5\sigma$ exclusion of the background-only model.

We adopt the standard LHC approach to discoveries~\cite{Cowan:2010js,ParticleDataGroup:2024cfk}, which is based on the test statistics variable defined in Eq.~(\ref{eq:tst}) evaluated at $G=0$
\begin{equation}\label{eq:tstD}
t_0(\o)=2\,\log\frac{\mathfrak{L}[\widehat{G},\widehat{\nu};\o]}{\mathfrak{L}[0,\hathat{\nu}_0;\o]}
=2\,\log\frac{\mathfrak{L}[\widehat{G},\widehat{\nu};\o]}{\mathfrak{L}[0,\bar\nu;\o]}
\,.
\end{equation}   
In the numerator likelihood, $\widehat{G}$ and $\widehat{\nu}$ are the values of the parameters that maximize the likelihood given the observed data. In the denominator, the likelihood is maximized over $\nu$ for the fixed value of $G=0$ that we are interested in testing. However, for $G=0$ the dependence of the theoretical predictions on the nuisance parameters drops out and the maximum, $\hathat{\nu}_0$, is at the maximum of the prior likelihood $\hathat{\nu}_0=\bar\nu$. 

\begin{figure}
    \centering
    \includegraphics[width=\linewidth]{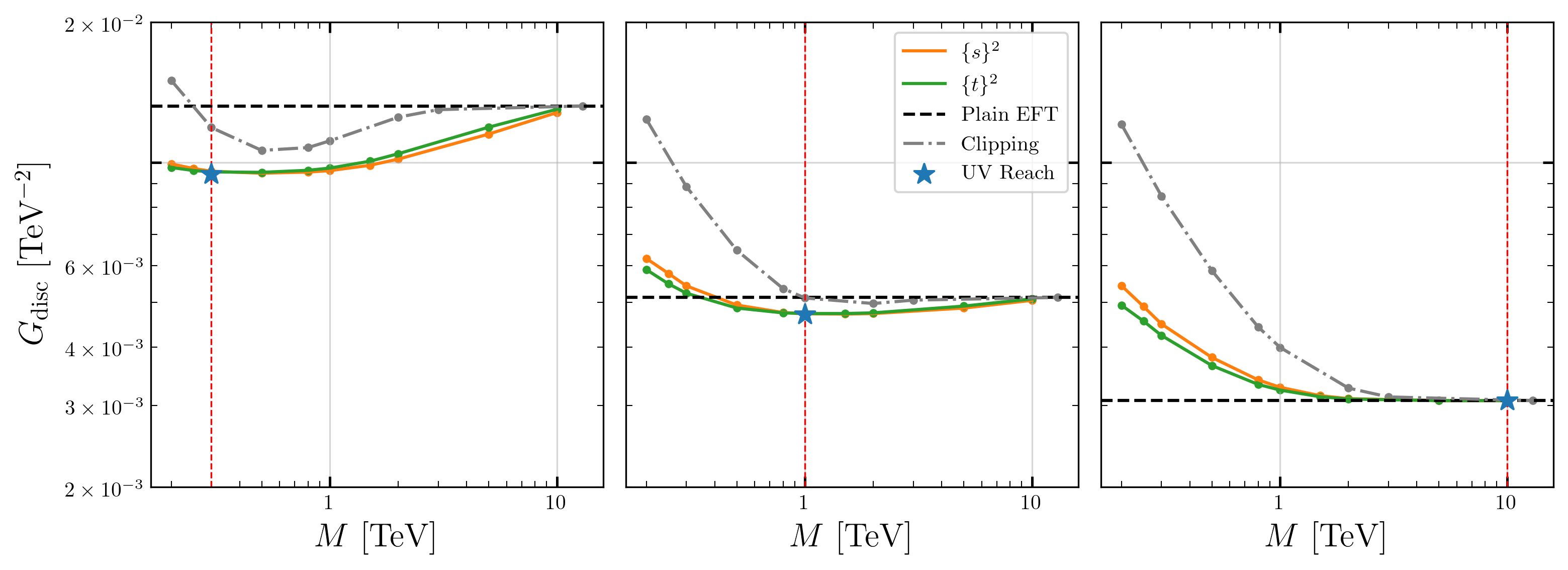}
    \caption{Discovery reach of the $m_{\ell\ell}$ analysis as a function of the cutoff scale $M$. The assumed true data distribution is the one predicted by the $t$-channel model, with $M_\phi$ equal to 300~GeV (left panel),  1~TeV (middle) and  10~TeV (right). The true resonance mass $M_\phi$ is reported as a vertical dashed line. The star is the reach of a dedicated UV model search.}
    \label{fig:discovery_reach}
\end{figure}

Using the large-sample and Asimov approximations discussed in \S\ref{ssec:SI} we can estimate $G_{\textrm{dis}}$ as follows. In the background-only hypothesis, $t_0$ is distributed as a $\chi^2$ with one degree of freedom, which is set to $25$, the threshold for a $5\sigma$ exclusion. The median of $t_0$, when the data follow a different model, can be estimated by setting the observed counts to the expected counts predicted by the model. The discovery reach is thus the solution to the equation
\begin{equation}
t_0(\e(G_{\textrm{dis}}))=25\,,
\end{equation}
where $\e(G)$ denotes the predictions of the $t$-channel model at fixed mass, after expressing the coupling in terms of $G$ as previously explained. The result is shown in Fig.~\ref{fig:discovery_reach} for the $m_{\ell\ell}$ binning using two nuisance parameters of the $s$-channel (orange line) or of the $t$-channel type (green line).

In Fig.~\ref{fig:discovery_reach}, the discovery reach of our method is compared with the plain EFT reach and with the reach of an EFT analysis that employs data clipping using the dilepton invariant mass as clipping variable, as explained in \S\ref{subExcDC}. The three panels reflect choices of the resonance mass of $M_\phi=$ 300 GeV, 1 TeV, and 10 TeV, as denoted by the vertical red dotted lines.  The discovery reach in these setups is estimated using the large-sample and Asimov approximations as previously explained. 
We see from the figure that our method gives a better discovery reach than the clipped EFT analysis, namely a smaller value of $G$ is sufficient for new physics discovery.
There are two effects that may help explain this.
First, the clipped EFT search treats the EFT prediction as perfect up to the scale $M_\text{clip}$, while we know that the EFT prediction has large uncertainties near the scale $M$;
second, the clipped search is completely throwing away the high-energy data above the scale $M_\text{clip}$, which may be useful for discovery even if they have non-negligible uncertainties.
The first problem can be ameliorated by lowering the scale $M_\text{clip}$, but this makes the second problem worse.
We believe that our approach is more accurately modeling the features of the underlying UV model, accounting for the improved discovery reach.
In support of this interpretation, we note that for the lighter two resonance masses, there is no choice of the clipping scale that can achieve the same discovery reach as our method.
Apparently, there is no choice of clipping scale that describes the data as well as our method when the cutoff $M$ is low enough for clipping to be relevant.

In Fig.~\ref{fig:discovery_reach}, we also compare to the discovery reach in a search for the UV model that actually described the data (indicated by a blue star).
Note that in all cases, the discovery reach for our method is very similar to the discovery reach for the UV model, provided that $M$ is chosen to be near the leptoquark mass $M_\phi$.
For values of $M$ far from $M_\phi$, the discovery reach weakens noticeably. 
We believe that this reflects the fact that for $M = M_\phi$, our method is doing a good job of mimicking the main features of the data from the UV model.
Of course, in an actual experimental search, the value of $M_\phi$ is not known, but the dependence on the reach as a function of $M$ may be useful to help characterize the discovery.

These results are very encouraging for our method, but the present analysis ignores several important effects.
First, our discovery reach on $G$ is based on a `local' $5\sigma$ significance, because the parameter $M$ 
has a fixed 
value throughout the analysis. 
The reduced sensitivity due to the `look-elsewhere'
effect should be estimated for a fair comparison with the plain EFT sensitivity,
which has no additional parameter.
Note that this caveat does not apply to the comparison with the plain EFT. 
Second, our results 
are based on the Asimov and large-sample approximations, which we quantified in Appendix~\ref{subcorr} 
for  
exclusion but not for a discovery analysis. Given the relatively small differences in $G_{\rm{disc}}$ observed, 
it is possible that 
deviations from 
these approximations may 
alter some of our conclusions. 
Finally, if the UV model has an $s$-channel resonance, we expect that the EFT approach will be much less effective than a targeted search for both exclusion and discovery, highlighting the complementarity between general EFT searches and searches for specific models.
Addressing these questions quantitatively would be needed for a fully conclusive assessment of our proposed method and is left to future work.

\section{Conclusions and Open Questions}\label{sec:conc}
We have presented a proposal to incorporate the theoretical uncertainties of an EFT into 
an experimental search with a well-defined statistical methodology.
We have focused on collider physics at the LHC, but the general ideas are much more widely applicable.
The challenge in constraining an EFT with experimental data is the fact that the theory is incomplete: its predictions at low energy are essentially an expansion in (positive) powers of $E/M$, where $E$ is a characteristic physical energy scale and $M$ is the energy scale where the EFT breaks down.
The full EFT contains infinitely many parameters and must be truncated to be predictive.
Regardless of the truncation, for $E \gsim M$ the EFT loses all predictivity.
If $M$ is sufficiently large, simple qualitative considerations can lead us to neglect the uncertainties due to the incompleteness of the EFT, but in many situations we need to quantify them in a statistical analysis.
This is mandatory at the LHC, where many EFTs that are being probed have a physical cutoff $M$ in the range of a few TeV, an energy scale where the LHC has significant parton luminosity.

Our proposal incorporates the theory uncertainties into the EFT predictions using nuisance parameters that modify the EFT predictions.
This is a standard technique in statistical analysis that is widely used at the LHC.
At low energies, the modified EFT predictions incorporate a subset of the higher-order EFT effects that mimic the most important corrections for the  process of interest.
Specifically, we propose parameterizing these corrections 
using the `Mandelstam descendants', higher-derivative terms arising from the expansion of heavy particle propagators in powers of momenta.
At higher energies where the EFT breaks down, the modified predictions should 
in principle parameterize the full range of allowed behaviors for the observable of interest.
However, an EFT is most useful for excluding a wide range of 
(possibly unknown) UV models and for this case the most important
requirement is that the modified EFT allows
the behavior at high energies to revert to that of the default
hypothesis (in our case, the Standard Model of particle physics).
This allows the EFT to mimic the default hypothesis for high-energy 
data, resulting in a conservative exclusion bound. 

This method requires as input prior information about the nuisance parameters, corresponding to an assumption about the prior probability (or prior likelihood) for the size of the EFT parameters that are 
neglected in the truncation. 
Assumptions about theoretical priors 
are often not made explicit in elementary particle physics (unlike other fields such as cosmology and astrophysics), but are  
commonly employed for the treatment of uncertainties of theoretical origin, such as the neglect of higher order radiative corrections.
Such priors are unavoidable for any theory prediction.
For a truncated EFT, this is particularly clear because it has no predictive power unless the size of the higher order terms is limited in some way.
In the context of LHC physics, we 
proposed a simple Gaussian prior that makes precise that the 
theoretical input that the
Mandelstam  descendants that parameterize the higher-order corrections are expected to be parameterized by parameters of at most order one.

We have tested these ideas in simulated LHC data with a simple model,
and compared our method 
to underlying UV models as well as existing proposals for addressing the problem of EFT incompleteness.
We show that our method gives robust and conservative results which are consistent when compared with optimized searches for UV models. 
Our 
method can 
be easily implemented with event reweighting, 
and avoids the problems 
of other approaches.

This work leaves for the future
a number of important further applications
and open questions.
First and foremost, we believe that it is important to apply these  ideas to actual searches at the LHC to develop an analysis pipeline and compare the results to 
existing  
approaches.
We believe that our proposal is both practical and robust (at least for exclusion), and can facilitate a wide range of EFT searches for deviations  from the Standard Model  that have so far not been performed, at least partly because 
there was no robust way of incorporating the potentially large 
theory uncertainties.

Improvements of the methodology should be also investigated.  
Further 
tests of the robustness of the conclusions, especially for discovery, beyond the Asimov and large-sample approximations should be performed.   Also, studying in particular the possible benefits of a more refined treatment of the $E>M$ regime where the EFT breaks down. 
Flexible functional interpolators such as neural networks or Gaussian mixture models could in principle be employed to 
give a more realistic modeling of 
the lack of knowledge of the amplitude in this regime. Practical challenges are the computationally demanding evaluation of the cross section predictions when the dependence on the parameters is not polynomial---preventing an analytic parametrization like \Eq{exp}---and the corresponding cost of the model training.

Finally, we believe that it may be useful to explore the
extension of these ideas beyond LHC physics.
Effective theories are widely employed in many areas of physics and the methods we have proposed may be 
useful whenever the higher order EFT corrections are not fully negligible.

Prior assumptions about neglected EFT terms are also important
in theoretical studies of EFTs, for example positivity bounds
on EFT coefficients can be derived assuming the general principles of unitarity, locality, and causality \cite{Adams:2006sv,Arkani-Hamed:2020blm,deRham:2017avq,Bellazzini:2020cot,Sinha:2020win,Tolley:2020gtv,Caron-Huot:2020cmc,Caron-Huot:2021rmr,Henriksson:2021ymi,Ma:2023vgc,Dong:2024omo}.  These can be important for existing searches, such as anomalous quartic gauge couplings (see e.g.~\cite{Zhang:2018shp,Bi:2019phv}).   
We hope that this work stimulates further discussion of these
topics.

\section*{Acknowledgements}
We have many people to thank for insightful discussions during the
long germination of this work:  Tim Cohen,
John Conway, 
Nate Craig,
Gauthier Durieux,
Christophe Grojean,
Da Liu,
Xioachuan Lu,
Andrew Pilkington,
Francesco Riva,  Michael Schmitt, Will Shepherd, and Kevin Zhang.
S.C. is supported in part by the DOE under grant DE-SC0011640.
M.L. is supported in part by the DOE under grant DE-SC-0009999.
The work of S.C. and M.L. 
was performed in part at the Aspen Center for Physics, which is supported by National Science Foundation grant PHY-2210452.
T.M is partly supported by the Yan-Gui Talent Introduction Program (grant No. 118900M128), Chinese Academy of Sciences Pioneer Initiative `Talent Introduction Plan', the Fundamental Research Funds for the Central Universities, and National Natural Science Foundation of China Excellent Young Scientists Fund Program (Overseas).
This work is part of the doctoral thesis of F.M., within the framework of the Doctoral Program in
Physics of the Autonomous University of Barcelona. 
The work of F.M. is supported by the Joan Oró predoctoral grant program of the Department of Research and Universities of the Government of Catalonia and co-funded by the European Social Fund Plus (Project reference: 2024 FI-1 00715).
This work is also 
part of the R\&D\&i project PID2023-146686NB-C31, funded by MICIU/AEI/10.13039/501100011033/ and by ERDF/EU. 
It is also supported by the Departament de Recerca i Universitats from Generalitat de Catalunya to the Grup de Recerca 00649 (Codi: 2021 SGR 00649).
\appendix

\section{Three-Point Couplings}\label{app:TP}

In this appendix, we outline the procedure for parameterizing higher-order
EFT corrections for BSM 3-point couplings.
To illustrate the ideas in a simple context, we will consider a BSM contribution to the 3-point self-coupling of the Higgs boson and consider the process $t\bar{t} \to hh$, neglecting gauge interactions.
\footnote{This process can be considered a simpler version of gluon fusion production of di-Higgs.}
The relevant terms in the SM Lagrangian are
\[
\scr{L}_\text{SM} &= \tfrac 12 (\partial h)^2 - \tfrac 12 m_h^2 h^2
- \frac{g_{hhh}}{3} h^3 + \cdots
\nn
&\qquad{}
+ \bar{t} i \gamma^\mu \partial_\mu t -
\frac{y_t}{\sqrt{2}} \left(v + h \right) \bar{t} t + \cdots,
\]
where
\[
g_{hhh} = \frac{3 m_h^2}{2v}.
\]
We write the leading BSM contribution to the Higgs cubic coupling as
\[
\scr{L}_\text{BSM} = -\tfrac{1}{3} G_{hhh} h^3.
\]

One way to incorporate the higher order EFT corrections to the Higgs cubic coupling is to introduce a form factor for the BSM 3-point coupling 
that depends on nuisance parameters.
We will use a different approach, for reasons we explain below.
We 
use the fact that higher order corrections to the Higgs cubic in the EFT are equivalent to BSM contributions to 4-point couplings involving the Higgs.
To see this, we note that using integration by parts,
the most general higher derivative corrections to the cubic Higgs self-coupling can be written as
\[
\eql{LBSMhhh}
\scr{L}_\text{BSM} = 
-\tfrac{1}{3} G_{hhh} \Bigg[ &
h^3 + \frac{c_1}{M^2} h^2 \Box h
+ \frac{c_2}{M^4} h^2 \Box^2 h 
 + \frac{c_3}{M^4} h (\Box h)^2+ \cdots 
\bigg].
\]
Using this to compute the tree-level BSM contribution to the amplitude gives
\[
\scr{M}_\text{BSM}(t\bar{t} \rightarrow hh) 
= G_{hhh} \frac{\scr{M}_\text{SM}(t\bar{t} \rightarrow hh)}{g_{hhh}}
\bigg[ 
1 &- \tfrac{1}{3} c_1 \frac{s + 2m_h^2}{M^2} + \tfrac{1}{3} c_2 \frac{s^2 + 2 m_h^4}{M^4} \nn
&{} + \tfrac 13 c_3 \frac{(2s + m_h^2)m_h^2}{M^4}
+ \cdots
\bigg] \eql{tri_form}
\]
where $\scr{M}_\text{SM}(t\bar{t} \rightarrow hh)$ is the Standard Model amplitude with an $s$-channel Higgs and $g_{hhh}$ is the SM cubic Higgs coupling.

We see that the higher derivative terms effectively modify the 3-point coupling, as well as giving additional terms suppressed by powers of $s$.
We could 
eliminate the additional terms depending on $m_h^2$ by replacing $\Box \to \Box + m_h^2$ in \Eq{LBSMhhh}.
This would allow us to define an `off-shell' form factor with the replacement $x \to (\hat{s} - m_h^2)/M^2$, {\it etc\/}.%
\footnote{This procedure is also applicable for more complicated 3-point couplings involving
particles with spin. 
This is because if one of the legs is off-shell, no matter how the derivatives act on the three legs, the only possible Lorentz scalar variables that can be generated are $s$ and the mass squared, $m^2$, of the on-shell legs. So we can always shift the  derivative to remove the trivial mass factor.}
Instead, we will
use the fact that the higher derivative corrections to 3-point couplings can be eliminated by field redefinitions, trading them for modifications of interactions involving 4 or more fields.
This works for general 3-point couplings, which is most easily understood using the on-shell amplitude approach to classifying the interactions.

Let us illustrate the field redefinitions in this example.  Taking $h\to h+ G_{hhh}X$, the Lagrangian becomes
\[
{\cal L} \to {\cal L} + G_{hhh}X\, \frac{\delta S}{\delta h}+O(G_{hhh}^2) = {\cal L} -G_{hhh} X\, (\Box h +m_h^2 h +\frac{y_t}{\sqrt{2}}\bar{t}t)+O(G_{hhh}^2).
\]
We write the field redefinition with a factor of $G_{hhh}$  
so that we can work in an expansion in this coupling.  By choosing $X$ appropriately, one can cancel terms in $\cal L$ with a $\Box h$, but adding terms that are 
equivalent to the substitution 
\[
\Box h = -m_h^2 h - \frac{y_t}{\sqrt{2}} \bar{t} t - G_{hhh}h^2 + \cdots,
\]
effectively using the equation of motion of $h$ to simplify the Lagrangian at $O(G_{hhh})$.
For terms with
more than one power of $\Box$,  
we can do multiple field redefinitions, which again amount to successive applications of the equations of motion. (In the on-shell amplitude approach, the fact that we can eliminate the $\Box$ terms follows immediately from the on-shell conditions for the external legs.)
Performing these field redefinitions gives
\[
\!\!\!\!\!
\scr{L}_\text{BSM} &\rightarrow
-\tfrac 13 G_{hhh} \Bigg[
\left(1-c_1\frac{m_h^2}{M^2}+(c_2+c_3)\frac{m_h^4}{M^4}\right) h^3
\nn
&\qquad\qquad\quad{}
-\frac{y_t}{\sqrt{2}} \left(\frac{c_1}{M^2}-\frac{(c_2+2c_3) m_h^2}{M^4}\right) h^2 \bar{t}t-\frac{y_t c_2}{\sqrt{2}M^4}  h^2 \Box(\bar{t}t) \Bigg]+O(G_{hhh}^2) 
\eql{tri_field_redef}
\]
That is, the form factor for the trilinear Higgs interaction is equivalent to further modification
of the $h^3$ coupling, along with the addition of 4-point couplings
such as $h^2 \bar{t} t$ to the BSM Lagrangian.  There are also 5-point interactions which are not relevant for $t\bar{t}\to hh$ at tree-level and also other terms (e.g.~$h^4$) if one works to higher order in $G_{hhh}$.  Using \Eq{tri_field_redef} to calculate $t\bar{t}\to hh$, we get the same amplitude as in \Eq{tri_form}, which can be seen by writing powers of
$s$ times the Higgs propagator using the identities
\[
\frac{s}{s - m_h^2} = 1 + \frac{m_h^2}{s - m_h^2},
\qquad
\frac{s^2}{s - m_h^2} = s + m_h^2
+ \frac{ m_h^4}{s - m_h^2},
\qquad\ldots,
\]
explaining the leading and subleading energy growth in \Eq{tri_form}.  

Because higher-derivative corrections to the cubic couplings are equivalent to 4-point couplings that are also interesting to constrain, we advocate the use of this parameterization of the EFT to obtain experimental constraints.
For the present example of $\bar{t} t \to hh$,  
we only need the 
$h^2 \bar{t} t$ coupling.
We therefore consider the modified  
BSM Lagrangian
\[
\tilde{\scr{L}}_\text{BSM} = 
-\tfrac 13 G_{hhh} h^3 
+ \tfrac 12 G_{hhtt} h^2 \bar{t} t.
\]
In Appendix B, we discuss the generalization of our method to more than one EFT coupling, and those will have to be used in this case.
For the present discussion, we assume that we are only constraining the modification of the cubic coupling, so we treat $G_{hhtt}$ along with its Mandelstam descendants as nuisance parameters.
In the terminology introduced above, the quartic coupling $G_{hhtt}$ can be
thought of as a Mandelstam descendant of the cubic coupling, where $G_{hhtt}$ is expected to be of order 
$G_{hhtt}\sim y_t G_{hhh} / \sqrt{2} M^2$. 
We therefore define the modified EFT coupling
\[
G_{hhtt} \to \frac{y_t G_{hhh}}{\sqrt{2}\,M^2}  
\left( \frac{F(x_\text{max})}{x_\text{max}} \right)^{{\mathrm p}/2}
\sum_{\alpha, \beta, \gamma} \nu_{\alpha\beta\gamma} F(x)^\alpha F(y)^\beta F(z)^\gamma,
\]
where $\nu_{\alpha\beta\gamma} \sim 1$.
Since the 4-point amplitude from the $hh\bar{t}t$ coupling grows linearly with energy as $m_{\bar{t}t}$, we choose $\textrm{p}=1$ so that it is unitary at high energies.
Note that in this case, $\nu_{000}$ is not fixed to 1; instead, 
it is the leading nuisance parameter and is assigned a prior distribution along with the others.  Again, in order to determine the reweighting dependence on $G_{hhh}$ and $G_{hhtt}$, it is important to utilize the technique described for multiple operators in Appendix \ref{app:MO}.

The model as defined so far does not cancel all possible unitarity violation at high energies.  The $hh\bar{t}t$ coupling leads to unitarity violating processes, such as $t\bar{t} \to W^+_L W^-_L h$ and $t\bar{b} \to W^+_L Z_L^4$ which cannot be canceled  \cite{Chang:2022crb}.  To make these unitarity, the prefactor power must be changed to $\textrm{p} =4$.  Similarly, a nonstandard Higgs cubic coupling by itself gives rise to unitarity violation in five-point and higher processes involving longitudinal gauge bosons, such as $V_L V_L \to h V_L V_L$ and $V_L V_L \to V_L V_L V_L V_L$, where $V=W, Z$ \cite{Chang:2022crb}.
Restricting to 2-to-2 processes, there is no unphysical energy growth from the Higgs cubic coupling modification alone, so we expect that most experimental searches using just the form factors and nuisance parameters introduced so far will give consistent results.
For other cubic couplings (such as $hZZ$) a modification of the cubic coupling does lead to unitarity violation in 2-to-2 scattering diagrams, and we must introduce a form factor for the cubic coupling to avoid this.
To illustrate the introduction of form factors for trilinear couplings, we return to our present example.
We define the kinematic invariants for each leg by
\[
v_{1,2,3} = \frac{p_{1,2,3}^2}{M^2},
\]
which can now be offshell, and then we can soften the behavior of the cubic coupling by replacing 
\[
G_{hhh}  \to G_{hhh} \left( \frac{F(v_\text{max})}{v_\text{max}} \right)^{\!\frac{\mathrm q}{2}},
\qquad
v_\text{max} = \text{max}\{ v_1, v_2, v_3 \}.
\]
Note we are not including a polynomial expansion in the $F(v_{1,2,3})$ because the higher-derivative corrections at low energies are taken into account by the nuisance parameters in the $hh\bar{t}t$ coupling.
We assume that $M > m_h$, so the rescaling leaves $G_{hhh}$ invariant at low virtuality.
At high virtuality the form factor scales as $\sim (M/E)^{\mathrm q}$, where $E^2 = \max\{ p_1^2, p_2^2, p_3^2\}$.
If we want to ensure that the five and six-point processes discussed above are unitary, one must require that $\text{q}\ge 2.$ 
In accordance with the discussion in \S\ref{ss:rFF}, we expect that the modified EFT amplitude will perform well for exclusion provided that it correctly parameterizes the higher derivative corrections at low energies and is able to reproduce the SM predictions at high energies.
This will be the case for the modified amplitude above for all values of $\text{p}$ and $\text{q}$ for which the full BSM contribution to the 2-to-2 scattering amplitude at high energies has the same (or faster) energy fall off as the SM.  
Because the modified $hh\bar{t}t$ coupling already has nuisance parameters that allow it to reproduce the SM, and larger values of $\text{q}$ in the form factor for the cubic further suppress its contribution to high energy events, we expect that the inclusion of a form factor for the cubic will not affect the results for exclusion. 
As discussed in \S\ref{subDisc}, the situation for discovery is more complicated, since in that case the optimal modification of the EFT is the one that reproduces the correct UV physics, which is unknown.

As can be seen from this discussion, the application of our method to the case of 3-point couplings has a number of additional complications and subtle points.
We leave it for future work to fill in the details, test the performance of the method, and provide a `how-to manual' for applying the method in concrete LHC searches.

\section{Multi-Operator Generalization}\label{app:MO}

In an analysis where several EFT operators are considered simultaneously, the formulas of \S\ref{sec:EGR} admit a straightforward generalization, which we report here for completeness. 

Following the notation used in the main paper, we assume that we are constraining a truncated EFT with operators $\cal O$ with Wilson coefficients $G_{\cal O}$.
The dependence of the event weights on the Wilson coefficients is a simple generalization of Eq.~(\ref{eq:rwg1})
\begin{equation}\label{eq:rwg11}
w_e(G)=w_e^{\textsc{sm}}\left[
1+\sum\limits_{\cal O} {\mathfrak{l}}^{\cal O}_e\,G_{\cal O}+\sum\limits_{{\cal O},{\cal O}'}{\mathfrak{q}}^{{\cal O},{\cal O}'}_{e}\,G_{\cal O} G_{{\cal O}'}
\right]\,,
\end{equation}
where the linear and quadratic coefficients ${\mathfrak{l}}$ and ${\mathfrak{q}}$ have been promoted to a vector and a symmetric tensor, respectively. 

The dependence on the nuisance parameters is introduced by a rescaling of the Wilson coefficients as in Eq.~(\ref{eq:Rdefn}). One set of independent nuisance parameters, $\nu^{\cal O}$, has to be introduced for each operator and the weights for the BSM contribution to the amplitude are
\[
\label{eq:rwg21}
w_e^{\textsc{eft}}(G,\nu)=w_e^{\textsc{sm}}\bigg[
&1+
\sum\limits_{\cal O}
{\mathfrak{l}}^{\cal O}_{e}\,\scr{R}_{\nu_{\cal O}}(x_1, x_2, x_3)\,G_{\cal O}\\
&\ \ +
\sum\limits_{{\cal O},{\cal O}'}
{\mathfrak{q}}^{{\cal O},{\cal O}'}_{e}\,
\scr{R}_{\nu_{\cal O}}(x_1, x_2, x_3)
\scr{R}_{\nu_{{\cal O}'}}(x_1, x_2, x_3)
G_{\cal O} G_{{\cal O}'}
\bigg] \,.\nonumber
\]

In the baseline setup with a linear dependence on the nuisance parameters of the amplitude rescaling factor, as in Eq.~(\ref{eq:FFonlyR}), the expected number of events in each bin takes the polynomial form 
\[
\label{eq:exp1}
\e_b(G,\nu)=
\e_b^{\textsc{sm}}\Bigg[ 
1 &+
\sum_{\cal O} \sum_{\alpha,\beta,\gamma}
G_{\cal O}\,(l^{\cal O}_{b})_{\alpha\beta\gamma}\nu^{\cal O}_{\alpha\beta\gamma}
\nn
&{}+ \sum_{{\cal O},{\cal O}'}
\sum_{\substack{\alpha,\beta,\gamma\\\alpha',\beta',\gamma'}}
G_{\cal O} G_{{\cal O}'}
(q^{{\cal O},{\cal O}'}_{b})_{\alpha\beta\gamma}^{\alpha'\beta'\gamma'}
\nu^{\cal O}_{\alpha\beta\gamma}\nu^{{\cal O}'}_{\alpha'\beta'\gamma'}
\Bigg]\,,
\]
where $\e_b^{\textsc{sm}}$ is the expected number of events in the background-only model. It should be noted that 
$\nu^{\cal O}_{000}$ are not a nuisance parameters; they are set to $\nu^{\cal O}_{000}=1$. The polynomial coefficients $l_b^{\cal O}$ and $q_b^{{\cal O},{\cal O}'}$ are given by the following sums over the events in the bin:
\[
(l_{b}^{\cal O})_{ \alpha\beta\gamma}=\frac1{\sigma^{\textsc{sm}}_b}\sum\limits_{e\,\in\, b}
w_e^{\textsc{sm}}{\mathfrak{l}}_e^{\cal O}
& 
\!\left( \frac{F({x_{\text{max},e}})}{x_{\text{max},e}}\right)^{\!\frac{\textrm{p}_{\cal O}}{2}}
F({x_{1,e}})^\alpha
F({x_{2,e}})^\beta
F({x_{3,e}})^\gamma,
\\[6pt]
(q_{b}^{{\cal O},{\cal O}'})_{\alpha\beta\gamma}^{\alpha'\beta'\gamma'}=
\frac1{\sigma^{\textsc{sm}}_b}\sum\limits_{e\,\in\, b}
w_e^{\textsc{sm}}{\mathfrak{q}}_e^{{\cal O},{\cal O}'}
& \! \left(\frac{F({x_{\text{max},e}})}{x_{\text{max},e}}\right)^{\!\frac{\textrm{p}_{\cal O} + \textrm{p}_{{\cal O}'}}{2}}
\nn
&{} \times
F({x_{1,e}})^{\alpha+\alpha'}
F({x_{2,e}})^{\beta+\beta'}
F({x_{3,e}})^{\gamma+\gamma'}.
\]
Here ${\textrm{p}}_{\cal O}$ are the powers defined for a single operator $\scr{O}$ below \Eq{stu123}.

\section{Corrections to the Large-Sample and Asimov Approximations}\label{subcorr}

Our results are based on a simplified statistical treatment---introduced in \S\ref{ssec:SI}---that relies on two approximations: the large-sample and the Asimov approximation. These approximations are well-established in LHC statistical practice, but they are subject to corrections that need to be quantified~\cite{Cowan:2010js}. The large-sample ($\chi^2$) approximation for the test statistics distribution in the `Null' hypothesis that we aim at excluding---i.e., in the EFT, for the exclusion setup of Eq.~(\ref{eq:tst})---is predicted by the Wilks theorem~\cite{Wilks:1938dza}. In turn, the theorem relies on the assumption that the data enable an accurate enough determination of the parameters such that the likelihood can be expanded around the true value of the parameters and treated approximately as Gaussian. This happens asymptotically for infinitely large data statistics, but how many data are concretely needed for the Gaussian approximation to work depends on the specific problem under study. The Asimov approximation is to estimate the median of the test statistics in the `Alternative' hypothesis that truly underlies the observed data---i.e., in the background-only SM hypothesis, for the exclusion setup of Eq.~(\ref{eq:tst})---by evaluating the test statistics on observed data that are equal to the predictions of the Alternative hypothesis. As far as we know, the foundations of the Asimov approximation are not rooted in a theorem, however its heuristic justification relies once again on a smooth nearly Gaussian likelihood.

It is legitimate to expect significant departures from the large-sample and Asimov approximation in our analysis, especially when the number of nuisance parameters is large. In fact, many nuisance parameters offer a redundant fit to the observed data and therefore we do not expect a single global maximum around which the likelihood can be expanded and approximated with a Gaussian, but rather a family of nearly degenerate maxima. On the other hand, precisely because of this redundancy, adding more and more nuisance parameters should not change the result and using a few nuisance parameters will be enough in practice. In this Appendix we assess the validity of the approximations using 2 nuisance parameters. Using more parameters is more demanding computationally and beyond the reach of our simple numerical implementation.

\begin{figure}
    \centering
    \includegraphics[width=1\linewidth]{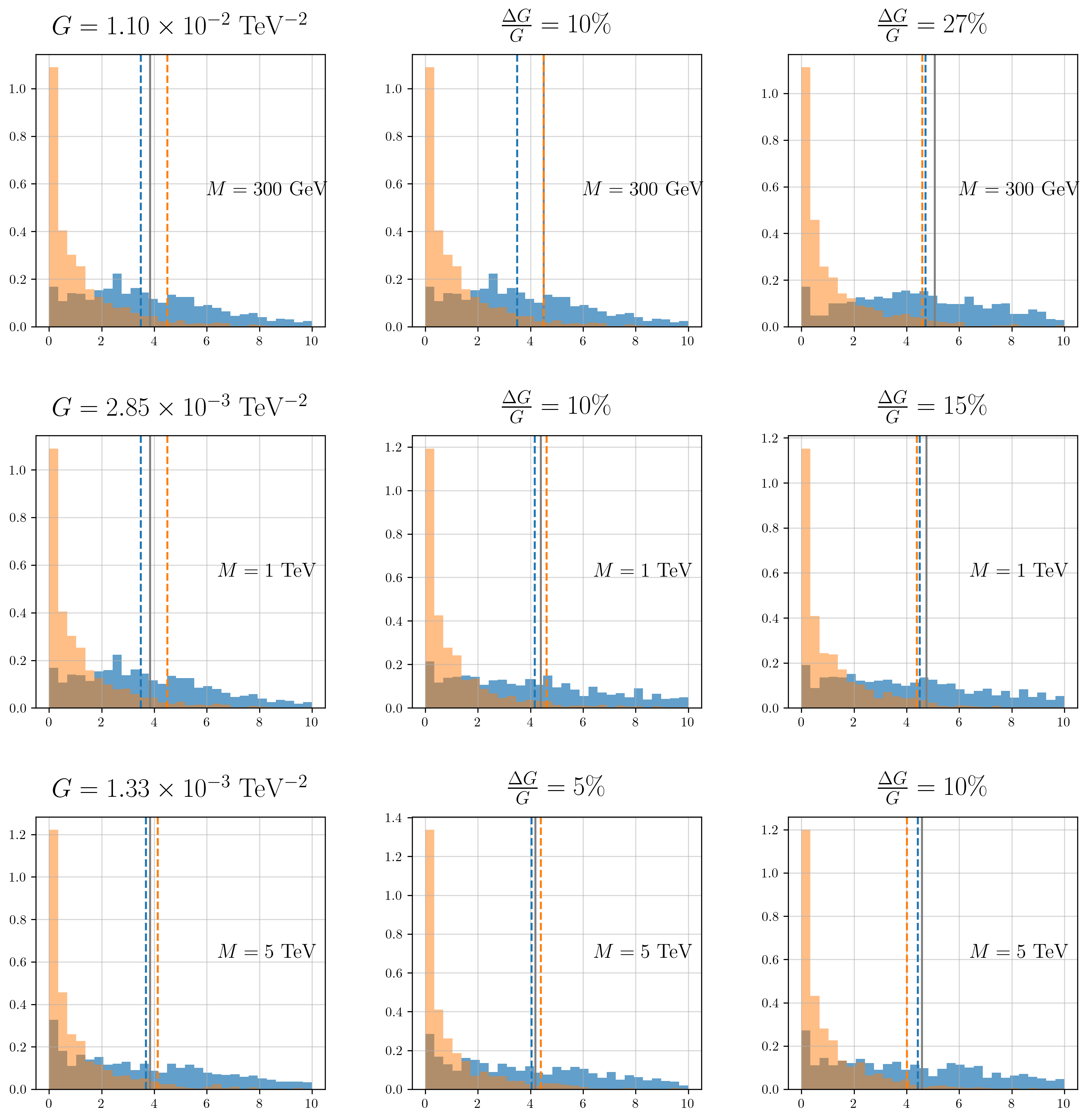}
    \caption{Distribution of the test statistics $t_G(\o)$~(\ref{eq:tst}) on toy datasets sampled from the SM (blue) and from the EFT (orange) hypotheses. The median of the former distribution and the $95\%$ probability threshold for the latter distribution are reported as vertical lines. The values of $M$ and $G$ are reported in each plot.}
    \label{fig:asimov}
\end{figure}

Going beyond the large-sample and the Asimov approximations requires generating toy instances of the observed counts in the bins, $\o_b$, and computing the test statistics $t_G(\o)$ for each toy dataset using Eq.~(\ref{eq:tst}). Toy datasets in the Alternative hypothesis---by which we assess the validity of the Asimov approximation computing the median of $t_G$---are straightforward to obtain because the Alternative hypothesis is sharply defined to be the SM hypothesis $G=0$. The toy datasets are generated by sampling $\o_b$ around the background-only expectation with independent Poisson distributions for each bin. By evaluating the test statistics $t_G(\o)$ for each toy dataset, we obtain the blue histograms displayed in Fig.~\ref{fig:asimov}. The median of each histogram is represented as a blue dashed vertical line, while the gray line---reported only on the left plot in each line---is the Asimov estimate of the median defined in Eq.~(\ref{tstAs}).

In Fig.~\ref{fig:asimov} we consider the $p_T$-binning setup, with $2$ nuisance parameters of the $s$-channel type. The corrections to the Asimov approximation are moderate. On the 3 lines, a low (300~GeV), medium (1~TeV) and high (5~TeV) value is considered for the cutoff scale $M$. The corrections to the Asimov approximation become smaller when $M$ is larger. This is because when $M$ is large the nuisance parameters have a small effect and the only parameter that is effectively left is the Wilson coefficient $G$. Since $G$ is precisely bounded by data, we do expect the Asimov approximation to work in this regime like it does in the plain EFT analysis.

The value of $G$ that is considered in the test statistics variable, $t_G(\o)$, is different in each plot. On the left plot of each line, it is set to the upper $2\sigma$ exclusion limit---see Fig.~\ref{fig:3binnings}---as estimated with the Asimov and large-sample approximation for the value of $M$ under consideration. With this choice of $G$, the Asimov estimate for the median of $t_G$ (gray line) is equal to $3.84$ on each plot, because this is the value that corresponds to the threshold for exclusion when assuming the large-sample ($\chi^2$) approximation for the distribution of the test statistics in the Null hypothesis. The actual median of $t_G$ (dashed blue line) is 
clearly  
different from $3.84$,
signaling corrections to the Asimov approximation. In the plots reported in the middle and in the right panels, the value of $G$ is progressively increased and the median consistently increases.

We turn now to an assessment of the validity of the large-sample approximation. In order to proceed, toy datasets need to be generated, following this time not the background-only but the Null (EFT) hypothesis with a specified value of $G\neq0$. This is not straightforward because in our setup the EFT predictions, $\e_b(G,\nu)$, depend on the value of the nuisance parameters and we have no criterion to select a specific point of the nuisance parameters space. The issue is not specific to our method, but on the contrary, generic of the profile-likelihood frequentist treatment of nuisance parameters. The fully consistent but impractical solution would be to repeat the toy generation for all values of $\nu$ and consider the envelope of the $t_G$ distributions obtained with the different values. In this way the threshold for exclusion is set to the highest values that is compatible with the Null at the given confidence level---for instance, $95\%$---for all values of $\nu$. The standard viable alternative is a `Bayesian' approach~\cite{ParticleDataGroup:2024cfk} in which $\nu$ is treated as a random variable and sampled from its prior distribution. For each sampled $\nu$, the expected $\e_b(G,\nu)$ are computed and one instance of the observed $\o_b$ in each bin is sampled from independent Poisson distributions. The distribution of $t_G(\o)$ obtained with this sampling strategy and Gaussian prior with $\sigma_\nu=3$, compatibly with Eq.~(\ref{Lik}), is displayed as orange histograms in Fig.~\ref{fig:asimov}. 

Owing to the large-sample approximation, the distribution of the test statistics in the Null hypothesis---orange histograms in Fig.~\ref{fig:asimov}---should match the $\chi^2$ distribution with one degree of freedom. Correspondingly, the threshold for exclusion at $95\%$~CL would be set at $3.84$, because this value defines the $5\%$ upper tail of the $\chi^2$ distribution with one degree of freedom. The true threshold for exclusion is obtained by computing the $5\%$ probability upper tail---i.e., the $95\%$ cumulative probability threshold---of the true distribution, and it is marked as a dashed orange vertical line in the plots. The departure from $3.84$ of the true threshold quantifies the corrections to the large-sample approximation.

The corrections to the large-sample and Asimov approximations that we observe in Fig.~\ref{fig:asimov} have a moderate impact on the the exclusion reach $G_{\textrm{exc}}$. Starting from the estimate of $G_{\textrm{exc}}$ obtained with the large-sample and Asimov approximation, which is considered in the plots on the left, we progressively raise $G$ in the middle and right plots until  the median of $t_G$ in the SM hypothesis (blue dashed) becomes larger than the $95\%$~CL threshold (orange dashed). The exclusion reach is the value of $G$ for which these two quantities coincide. From the results shown on the left column of the figure we can thus conclude, for instance, that the exclusion reach is less than $25\%$ weaker than the estimate when $M=300$~GeV. The corrections to $G_{\textrm{exc}}$ are even smaller for larger $M$.

\bibliographystyle{JHEPAW}
\bibliography{biblio}

\end{document}